\documentclass[11pt,singlecolumn,aip,jcp,groupedaddress]{revtex4-1}
\usepackage[colorlinks, linkcolor = black, citecolor = black, filecolor = black, urlcolor = blue]{hyperref}

\usepackage{graphicx}
\usepackage{subcaption}
\usepackage{amsmath}
\usepackage{amssymb}
\usepackage{float}
\usepackage{amsmath}
\usepackage{bm}
\usepackage{fixmath}
\usepackage{framed,color}
\usepackage[makeroom]{cancel}
\definecolor{shadecolor}{rgb}{0.85,0.85,0.85}

\DeclareGraphicsExtensions{.pdf,.gif,.jpg}

\newcommand{\bra}[1]{\ensuremath{\left\langle#1\right|}}
\newcommand{\ket}[1]{\ensuremath{\left|#1\right\rangle}}

\begin{document}
\title{Chebyshev Hierarchical Equations of Motion for Systems with Arbitrary Spectral Densities and Temperatures }
\author{Hasan Rahman} 
\affiliation{Department of Physics and Earth Sciences, Jacobs University Bremen,
Campus Ring 1, 28759 Bremen, Germany}
\author{Ulrich Kleinekath\"ofer}
\email{u.kleinekathoefer@jacobs-university.de}
\affiliation{Department of Physics and Earth Sciences, Jacobs University Bremen,
		Campus Ring 1, 28759 Bremen, Germany}

\begin{abstract}

The time evolution in open quantum systems, such as a molecular aggregate in
contact with a  thermal bath, still poses a complex and challenging problem. The
influence of the thermal noise can be treated using a plethora of schemes,
several of which decompose the corresponding correlation functions in terms of
weighted sums of exponential functions. One such scheme is based on the
hierarchical equations of motion (HEOM), which is built using only certain forms
of bath correlation functions. In the case where the environment is described by
a complex spectral density or is at a very low temperature, approaches utilizing
the exponential decomposition become very inefficient. Here we utilize an
alternative decomposition scheme for the bath correlation function based on
Chebyshev polynomials and Bessel functions to derive a hierarchical equations of
motion approach up to an arbitrary order in the environmental coupling. These
hierarchical equations are similar in structure to the popular exponential HEOM
scheme, but are formulated using the derivatives of the Bessel functions. The
proposed scheme is tested up to the fourth order in perturbation theory for a
two-level system and compared to benchmark calculations for the case of
zero-temperature quantum Ohmic and super-Ohmic noise. Furthermore, the benefits
and shortcomings of the present Chebyshev-based hierarchical equations are
discussed.

\end{abstract}
	
\maketitle

\section{Introduction}

An accurate description of quantum dissipation continues to be formidable
challenge. While various reduced density matrix (RDM) based
schemes~\cite{weis99,breu02,may11} can account for the external influences,
e.g., from a protein environment, on the excitation energy transfer in systems
such as pigment-protein complexes, there exist only a few methods that can
tackle arbitrary intra-system and system-environment couplings
\cite{tani89,ishi09a,prio10a,nalb11a,muehl12a,schr15c}. One  way to determine  the RDM
dynamics is via quantum master equations (QMEs). The latter methods, however,
can usually only account for  the lowest orders in the system-bath interaction
due to their complex and very tedious extension to higher orders. This
limitation is not present in the well-known hierarchical equations of motion
(HEOM) approach \cite{tani89, xu07a, xu04b, tani06, yan04, shao04a} first
formulated by Tanimura and Kubo~\cite{tani89} for bosonic environments. The HEOM
scheme employs a possibly infinite hierarchy of auxiliary density operators
(ADO) to generate arbitrary orders of system-bath coupling in a systematic and
iterative fashion. Thus, in principle the HEOM approach allows for an exact
treatment of the system-bath interaction and can be used to obtain exact
solutions for the energy transfer in systems such as light-harvesting complexes
for arbitrary exciton-phonon coupling strengths. The hierarchy has theoretically
an infinite number of terms and may be truncated at a finite level depending on
the strength of the system-environment coupling. For the case of low
temperatures \cite{ishi05a,amin09a}, certain approximations can render the HEOM
computationally highly efficient~\cite{ishi09b}. Although the HEOM scheme
remains one of the most powerful and well-known approaches, it is
computationally expensive for larger systems, and depending on the chosen
parameters, may require tens of gigabytes of computer memory and large amounts
of computation time. GPU implementations \cite{krei11a} of the HEOM provide a
considerable improvement in computation time, but due to GPU memory limitations
the approach is usually constricted to systems composed of only a few
chromophores. HEOM implementations on parallel computers \cite{stru12b} and/or 
in a distributed memory fashion might render the scheme applicable to larger
systems \cite{kram18a}. Furthermore, the HEOM may be reformulated in terms of a linear transformation to gain a considerable computational advantage~\cite{wilk15a}.

The standard HEOM equations can only be constructed for  specific forms of the
environment correlation functions. Typically the hierarchical linkages are built
using an exponential form of the correlation function, i.e., the correlation
functions are represented as  weighted sums of exponential functions. This
scheme is  also termed exponential decomposition approach
\cite{tani06,stru12b,stru12c}. However, using the exponential decomposition
implies that the corresponding spectral densities of the environment are
generally constricted to certain specific forms such as the one composed of
Lorentzian functions and the Bose-Einstein function is approximated by a series
expansion such as the Matsubara series. This fitting procedure of a given
spectral density in terms of Lorentzian functions can be complicated in
practice~\cite{liu14a} while numerical techniques such as  least-squares solver
may be employed to minimize the error of the Lorentzian reconstruction. The
Ohmic spectral density, for instance, can be decomposed into a sum of Lorentzian
functions \cite{meie99}. For realistic spectral densities such as those
extracted from atomistic simulations, the spectral density can exhibit a highly
nonlinear behavior  and cannot be represented accurately using a finite sum
Lorentzian functions. To overcome this obstacle, schemes to fit certain  classes
of functions better suited to mimic super-Ohmic behavior do exist \cite{rits14a}.
However, the complications regarding the fitting procedures only escalate
especially when reconstructing experimentally or numerically obtained spectral
densities, which, at least for certain systems, show sharply varying energy
profiles \cite{chan15a,jang18a}. Nonetheless, a wide variety of spectral densities may be dealt with using the scheme presented in Ref.~\citenum{datt12a}. %

As an alternative to the regular exponential decomposition method and the
corresponding Lorentzian expansion of the spectral density, an extended version
of the HEOM scheme was presented by Tang and co-workers \cite{tang15a} where a
complete set of orthonormal functions is introduced to expand the bath
correlation function. The HEOM is then constructed using auxiliary fields
composed of these orthonormal functions. This expansion of the correlation
function allows the extended HEOM to be usable for the zero-temperature case
where the original exponential HEOM version faces an extreme numerical
inefficiency. It is also possible, however, as we show in this work, that a
hierarchically linked set of equations be built using the Chebyshev
decomposition scheme~\cite{pope15a, pope16a, rahm18a}. If the bath correlation
function is expanded in terms of Chebyshev polynomials and Bessel functions
instead of exponentially decaying functions, the derivatives of Bessel functions
allow the construction of a set of hierarchical equations which are similar in
structure to the exponential decomposition based HEOM. This Chebyshev decomposed
HEOM scheme (C-HEOM) has the clear advantage that no expansion  of the spectral
density of the environment is necessary and any kind of environmental spectral
density can be incorporated into the scheme in an exact manner. Furthermore, all
temperatures can be simulated at the exact same numerical cost. %
The Chebyshev decomposition has been previously applied to the second-tier
truncation of the HEOM \cite{tian13a} for Fermionic systems to simulate
arbitrary temperatures. These equations correspond to the Keldysh nonequilibrium
Green’s functions scheme \cite{croy09a}, an approach that is exact for
noninteracting particles. The HEOM itself have been applied to a range of
Fermionic systems. An extension of the HEOM approach (RSHEOM)~\cite{erpe18a},
which uses an analytic re-summation over poles, also shows similar advantages
and drawbacks as the scheme presented in this work. In earlier
studies~\cite{pope15a, rahm18a}, we have reported Chebyshev decomposed QMEs as
well as a nonequilibrium Green's function scheme for Fermionic systems, which
present the advantage of not being limited to  simple forms of spectral
densities or being inefficient for low temperatures.%

Polynomial expansions such as the Chebyshev expansion employed in the present
study have been applied previously in the context of wave packet
propagation~\cite{kosl94}, for inhomogeneous~\cite{ndon09a} as well as for
explicitly time-dependent cases~\cite{ndon10a}. For open quantum systems
polynomial-based schemes such as those based on  Chebyshev polynomials
\cite{guo99} as well as on Faber and Newton polynomials \cite{huis99} have been
studied. Furthermore, a time-evolving density with orthogonal polynomials
algorithm (TEDOPA) has been reported in the literature that is able to treat
dissipative systems with arbitrary spectral densities \cite{wood14a}. Ensemble average methods such as the hierarchy of pure states HOPS~\cite{sues14a} scheme are also able to deal with complex spectral densities at stronger coupling strengths.
Another efficient method for the time propagation of the total wave function is
the multilayer multiconfiguration time-dependent Hartree (ML-MCTDH)
theory~\cite{wang03, wang08b}. The quantities of the system of interest, e.g.,
the reduced density matrix (RDM) of the system can be estimated by the partial
trace over the bath degrees of freedom.  Alternatively using path integral-based
methods~\cite{feyn65},  the time evolution of the RDM can then be calculated by,
e.g., the quasi-adiabatic propagator path integral (QUAPI)~\cite{nalb10a} or the
stochastic path integral \cite{moix12a}. 
%

The HEOM formalism may be derived using the path integral approach utilizing the
Feynman-Vernon influence functional formalism \cite{tani89, xu07a,xu04b,tani06},
or alternatively, by separating the system and bath degrees of freedom via
stochastic fields as shown by Shao et al.~\cite{yan04, shao04a}, leading to the
same equations of motion in both cases. While the latter approach dissociates
the system and bath degrees of freedom using stochastic fields in this work we
utilize the former. In the present study the HEOM approach based on the the
Chebyshev decomposition is derived from scratch utilizing the path integral
formalism and then applied to some test cases. To this end, the article is
organized as follows. In the subsequent section the typical tight-binding model
for open quantum systems is presented, while in section III the Chebyshev
expansion is summarized. In section IV we employ the Chebyshev expansion to
derive  hierarchically linked equations of motion and present their general
formulas. Numerical results for test examples are shown and discussed in section
V followed by a final conclusions. More specifically, we apply the newly
developed C-HEOM to a two-level model and compare the results with other
standard methods such as the exponential HEOM and ML-MCTDH.

\section{Model Hamiltonian}

In the theory of open quantum systems, the Hamiltonian of a composite system is
usually considered to be composed of three constituent parts,
\begin{equation}
\label{TotHam}
H(t) = H_{S} + H_{R} + H_{SR}~,
\end{equation}
i.e., the total Hamiltonian $H(t)$ is split into the relevant system
Hamiltonian, $H_{S}(t)$, for the molecular aggregate, and the thermal bath
Hamiltonian, $H_{R}$,  while $H_{SR}$ represents the system-reservoir coupling
Hamiltonian. The system Hamiltonian is assumed to be of the tight-binding type
and, for example, for light-harvesting complexes one usually restricts oneself
to the single-exciton manifold. In this case, the basis states of the system
$|i\rangle$ mimic potential states of single excitons. The ground state  without
any excitation, $|0\rangle$, is usually not considered in these models. In
general, the system is assumed to be composed of $N$ sites with site energies
$E_i$ at site $i$. Sites $i$ and $j$ interact via electronic couplings,
$V_{ij}$. Each site may thus be coupled to any other site. The Hamiltonian of
the $N$-site system in tight-binding description can be written as
\begin{align}
H_S=\sum_i^N E_i \ket{i}\bra{i} +\sum_{i \neq j}^N   V_{i j} \ket{i}\bra{j}~. 
\label{system_hamiltonian}
\end{align}
For the case of exciton transfer it is possible to calculate the exciton
energies and the electronic couplings using the electronic structure of the
system constituents. The resulting information may then be used to parameterize
the above mentioned tight-binding model  \cite{chan15a,jang18a}. Each thermal reservoir $\alpha$ with chemical potential $\mu_{\alpha}$ is described by the Bose-Einstein statistics
\begin{equation}
n_\alpha(\epsilon) = \frac{1}{ e^{(\epsilon - \mu_\alpha)\beta} -1}~.
\end{equation}
and is given as a sum of displaced harmonic oscillators.
 The system-reservoir coupling Hamiltonian 
$H_{SR}$ can be written as a sum of products of system, $K_j$, and bath
operators, $\Phi_j$. As for the application have excitonic systems in mind, the system operators are assumed to be site
diagonal for excitonic systems \cite{may11} 
\begin{equation}
H_\mathrm{{\it SR}}=\sum_j \Phi_j K_{j} = \sum_j \sum_\xi c_{j \xi} x_{\xi} 
\ket{j}\bra{j}~.
\label{h_coup}
\end{equation}
In the above equation, $ x_{\xi}$ denotes the displacements of the harmonic bath
coordinates $\xi$ and $c_{j \xi}$ the system-reservoir coupling strengths. The
operator $K_j  = \ket{j}\bra{j}$, i.\ e., the projector onto site $j$, described
the system part of the system-reservoir coupling. This choice insures that an
interaction between the system and the environment at site $j$ exists only if
the site hosts an excitation. This description is limited to situations in which
each site has an independent thermal bath that is uncorrelated to the bath modes
from other sites. The Chebyshev approach outlined below can be extended to other
types of system-bath interactions in a straightforward manner.

\section{Chebyshev Decomposition of the Correlation Functions}

In this section, we follow the line of reasoning presented in our studies
\cite{pope15a,pope16a,rahm18a} to expand the bath correlation functions in terms
of Chebyshev polynomials. To make use of  Chebyshev polynomials, a mapping of the
spectral range of the Hamiltonian to their interval of definition, $[-1,1]$, is
required. In a first step, the infinite integration range usually present in
correlation functions is restricted to a finite range $[\omega_{min},
\omega_{max}]$. Subsequently, by introducing  the dimensionless variable $x =
(\omega-\bar{\omega})/\Omega$, where $\bar{\omega} = (\omega_{max} +
\omega_{min})/{2}$ and $\Omega = (\omega_{max} - \omega_{min})/{2}$, the
integral expression for the correlation function can be written as
\begin{equation}
C_j(t) = \frac{\hbar \Omega}{\pi}  \int_{-1}^1 \mathrm{d}x~ \frac{e^{ -i(\Omega
x+\bar{\omega})t} \mathcal{J}_{j}(\Omega x + \bar{\omega})}{1-e^{-\beta \hbar
(\Omega x + \bar{\omega})}} ~. \label{corrplus}
\end{equation}
with $\mathcal{J}_{j}$ denoting the spectral density of the environment  which for the present scheme can assume any form. The correlation functions play a major role inside the auxiliary density operators (ADOs), $\Lambda_j(t)$, which appear inside the definition of the influence
functional given in below Eq.~\ref{inf_func} below and are defined as
\cite{klei04a}
\begin{equation}
\Lambda_j(t) = \int_{0}^{t} \mathrm{d}\tau~C_j(t-\tau)
U_S(t,\tau)K_jU_S^{\dagger}(t,\tau)~, \label{def_of_lambda}
\end{equation}
where $U_S(t,\tau)$ denotes  the time-evolution operator of the system. The
next step is to expand the exponential term inside the Fourier kernel in  Eq.~\ref{corrplus} by using the Jacobi-Anger identity \cite{arfk85} given by
\begin{equation}
e^{ -i \Omega x t} = J_0(\Omega t) + \sum_{k=1}^{\infty} 
2( 
-i)^k \mathnormal{J_k}(\Omega t) \mathnormal{T_k(x)}\ \ , \ \ \forall t \in \mathbb{R} \ , \ \forall x \in [-1,1]~,
\label{ja-identity}
\end{equation}
where $T_k$ and $J_k$ denote the Chebyshev polynomials and the Bessel functions of 
the first kind, respectively. Consequently, the correlation functions can be written as
\begin{equation}
C_j(t) = \sum\limits_{k=0}^{\infty} \underbrace{ \mathnormal{J_k}(\Omega t) e^{
-i \bar{\omega} t}}_{C_k(t)} \underbrace{ \frac{\hbar\Omega}{\pi} \int_{-1}^{1}
\mathrm{d}x~ (2-\delta_{0,k})( -i)^k \mathnormal{T_k}(x) \frac{
\mathcal{J}_{j}(\Omega x + \bar{\omega})}{1-e^{-\beta \hbar (\Omega x +
\bar{\omega})}} }_{I_{j,k}} = \sum_{k=0}^{\infty}{C_k}(t) I_{j,k}~,
\label{def_of_CnandIn}
\end{equation}
where $\delta_{0,k}$ denotes the Kronecker delta function. Moreover, in
Eq.~(\ref{def_of_CnandIn}) the time-dependent quantities $C_{k}(t)$ as well as
the time-independent integrals $I_{j,k}$ are defined and grouped separately. The time-independent quantities $I_{j,k}$ need to be calculated once at the start of a simulation, while the time-dependent part of the correlation functions resides in the expression for $C_k(t)$.
Moreover, it is clear that no assumptions were made regarding the form of
spectral densities or on the Bose-Einstein distribution, hence on the involved
temperature. Thus, such an expansion is able to embed complex spectral densities
and handle low temperatures. The Chebyshev polynomials of higher orders, i.\ e., larger values of $k$, show a strongly oscillating behavior not amenable to standard techniques of numerical integration. To this end, the use of a property of the Chebyshev polynomials, $T_k(\mathrm{cos}(\Theta)) = \mathrm{cos}(k \Theta)$, renders the integrals, $I_{j,k}$, to the general form $f(x) \cdot \mathrm{cos}(\omega x)$, a form that can easily be dealt with using specifically designed integration subroutines from the QUADPACK library \cite{dedo83a}.

The central idea of the exponential decomposition used in formulating the QME or the HEOM is to decompose the correlation functions into a sum of exponential functions \cite{klei04a}, typically formulated in conjunction with a complex-pole expansion of the Bose-Einstein distribution. This allows for writing the time derivatives of the partial correlation functions in terms of themselves, enabling the formation of closed forms of equations. To achieve the latter goal one can avail an interesting, comparable property \cite{arfk85} of the derivatives of the Bessel functions of the first kind that appear inside the time-dependent part of the bath correlation function in Eq.~(\ref{def_of_CnandIn}):
\begin{equation}
\frac{\mathrm{d}}{\mathrm{d}t} \mathnormal{J}_k(\Omega t) = \frac{\Omega}{2}
\big(\mathnormal{J}_{k-1}(\Omega t) - \mathnormal{J}_{k+1}(\Omega t) \big)~.
\end{equation}
Since the derivatives of a certain order of Bessel functions can be written as a sum of Bessel functions of its nearest-order neighbours, the time derivative of ${C_k}(t)=\mathnormal{J_k}(\Omega t)
e^{-i\bar{\omega} t}$ can be written as
\begin{equation} 
\frac{\mathrm{d}}{\mathrm{d}t}{C_k}(t) = -i\bar{\omega} J_k(\Omega t) e^{-
i\bar{\omega} t} + e^{-i \bar{\omega} t} \frac{\mathrm{d}}{\mathrm{d}t}
\mathnormal{J}_k(\Omega t) = -i\bar{\omega}{C_k}(t) + \frac{\Omega}{2} \big(
{C}_{k-1}(t) - {C}_{k+1}(t) \big)~, \label{eq_deriv_C}
\end{equation}
where the derivative of ${C_k}(t)$ can be seen to be related to itself and its nearest-order
neighbours. This expression is the key point utilized in the derivation of the
C-HEOM.\\ %
%
%
%
Using the Chebyshev expansion one can rewrite Eq.~(\ref{def_of_lambda}) as
\begin{equation} 
\Lambda_{j}(t) = \sum_{k=0}^{\infty} I_{j,k} \underbrace{ \int_{0}^{t} d\tau~
{C}_k(t-\tau) U_s(t,\tau) K_j U_s^{\dagger}(t,\tau) }_{\Lambda_{j,k}(t)} =
\sum_{k=0}^{\infty} I_{j,k} \Lambda_{j,k}(t)~. \label{def_part_ado}
\end{equation}
Thus, the $\Lambda_{j}(t)$ are dissociated into  time-independent and
time-dependent parts, and the quantities $\Lambda_{j,k}(t)$ now carry the
time-dependence. Their defining expression, Eq.~\ref{def_part_ado}, can be
differentiated in a straightforward manner and used together with the relation
in Eq.~(\ref{eq_deriv_C}) to construct a hierarchically linked set of equations.
For practical purposes, the infinite sum over the partial ADOs,
$\Lambda_{j,k}(t)$, needs to be truncated at a finite order $K_{ch}$. As
detailed in Ref.~\citenum{pope15a} and similar to the method of
Ref.~\citenum{tian12a}, truncating at a certain $K_{ch}$ limits the approach to
a simulation time dependent on the truncation order $K_{ch}$.  The uniform
Chebyshev decomposition scheme shows an exponential convergence, generating
extremely accurate results given the criteria for convergence are fulfilled
\cite{kosl94}.


\section{Chebyshev Expansion Applied to HEOM}

In order to derive a HEOM version based on the Chebyshev decomposition, we start, in analogy to the derivation of the exponential HEOM \cite{schr15b} 
with the time evolution of the reduced density matrix, $\bm{\rho}_S(t)$,
obtained by applying the time evolution operator to the density matrix at $t_0$, 
\begin{eqnarray}
\bm{\rho}_S(t)= \text{tr}_{\text{bath}} \big\{ U(t,t_0) \bm{\rho}(0) U^{\dagger}(t,t_0)  \big\} = \tilde{\mathcal{U}}(t,t_0) \bm{\rho}_S(t_0)~.
\end{eqnarray} 
The Liouville space propagator, $\tilde{\mathcal{U}}(t,t_0)$, depends on the
system part as well as the system-bath interaction part of the total system. To
factorize the total system or to separate the system and bath DOFs, the
Feynman-Vernon influence functional formalism~\cite{feyn63a} needs to be
utilized by rewriting $\tilde{\mathcal{U}}(t,t_0)$ in the path integral formalism.
The time evolution operator $U(t,t_0) $ in path integral representation is given
by \cite{may11}
\begin{eqnarray}
U(t,t_0) = \int_{\alpha_0}^{\alpha_t} \mathcal{D}
\alpha~e^{\frac{i}{\hbar}S_{\rm total}[\alpha]}~,
\end{eqnarray} 
where $\alpha$ denotes arbitrary paths in phase space with the respective fixed
starting and ending points $\alpha_0$ and $\alpha_t$. The action functional
$S_{\rm total}[\alpha]$ describes the evolution of the total system but can be decomposed
into system, bath, and system-bath interaction parts. The time evolution of the
full density matrix in path integral formalism therefore reads~\cite{xu07a}
\begin{align}
\rho_{\alpha, \alpha', t} =& \int_{\alpha_0}^{\alpha_t} \mathcal{D} \alpha
\int_{\alpha'_0}^{\alpha'_t} \mathcal{D}
\alpha'~e^{\frac{i}{\hbar}(S_{S}[\alpha] +S_{B}[\alpha] + S_{SB}[\alpha] )}
\rho_{\alpha_0, \alpha'_0, t_0} e^{-\frac{i}{\hbar}(S_{S}[\alpha']
	+S_{B}[\alpha'] + S_{SB}[\alpha'] )}\\\nonumber & = \mathcal{U}(\alpha, \alpha',
t; {\alpha_0, \alpha'_0, t_0}) \rho_{\alpha_0, \alpha'_0, t_0}
\end{align} 
By tracing over the bath DOFs one gets~\cite{xu04b}
\begin{align}
\mathcal{U}(\alpha, \alpha', t; {\alpha_0, \alpha'_0, t_0}) =
\int_{\alpha_0}^{\alpha_t} \mathcal{D} \alpha \int_{\alpha'_0}^{\alpha'_t}
\mathcal{D} \alpha' ~e^{iS[\alpha]} \mathcal{F}(\alpha,\alpha') e^{S[\alpha']}~.
\label{time_evl_op}
\end{align} 
In this expression the Feynman-Vernon influence functional $
\mathcal{F}(\alpha,\alpha') $ represents the system-bath
interaction~\cite{feyn63a}. We note in passing that it is of course important to
maintain the order of the terms in Liouville space as, for instance, the time
evolution operator and the density operator do not commute. Employing the
Caldeira-Leggett model for the system-bath interaction and using a single system
operator $K$, the influence functional can be written as (adopting an arbitrary
Liouville space representation $\bm{\alpha} = (\alpha, \alpha')$)
\begin{eqnarray}
\label{inf_func}
\mathcal{F}[\bm{\alpha}(t)]= \text{exp} \Big\{- \int_{t_i}^{t} d\tau
K^{\times}[\bm{\alpha} (\tau)]~ \Lambda^{\times}[\bm{\alpha}(\tau)] \Big\}~.
\end{eqnarray} 
In the above, the  symbol ${\times}$ denotes a commutator, i.e.,
\begin{eqnarray}
\mathcal{F}[\bm{\alpha}(t)] = \text{exp}\Big\{-\sum_{j} \int_{t_i}^{t} d\tau (K_{j}[\bm{\alpha}(\tau)] - K_{j}^{\dagger}[\bm{\alpha}(\tau)])~ (\Lambda_j[\bm{\alpha}(\tau)]  - \Lambda_j^{\dagger}[\bm{\alpha}(\tau)] ) \Big\}~.  \label{inf_f1}
\end{eqnarray}

Taking the time derivative of $\bm{\rho}_S(t)$ leads to an equation of motion
describing the time evolution of the system in connection with the external
bath. Therefore, the derivative of the time evolution operator defined earlier,
$\mathcal{U}(\alpha_t, \alpha'_t, t; {\alpha_0, \alpha'_0, t_0}) $, needs to be
evaluated. This operator $\mathcal{U}(\alpha_t, \alpha'_t, t; {\alpha_0,
	\alpha'_0, t_0}) $ basically contains the influence functional
$\mathcal{F}(\alpha, \alpha')$ and the time derivatives of the action term
describing the evolution of the unperturbed system, given by
\begin{eqnarray}
\frac{\partial}{\partial t}~ e^{iS_{S}[\alpha]}= -i H_S~e^{iS_{S}[\alpha]}~,
\end{eqnarray} 
while the time derivative of influence functional, containing the influence of
the system-bath interaction, is given by
\begin{eqnarray}
\frac{\partial}{\partial t}~ \mathcal{F}[\bm{\alpha}(t)] = - K^{\times}[\alpha (t)]~  \Lambda^{\times}[\alpha (t)]~  \mathcal{F}[\bm{\alpha}(t)]  ~.
\end{eqnarray} 
To evaluate the above expression further, one needs to introduce the so-called
auxiliary influence functional (AIF) $\mathcal{F}_1$ so that
\begin{eqnarray}
\frac{\partial}{\partial t}~ \mathcal{F}[\bm{\alpha}(t)] = - K^{\times}[\alpha (t)]~   \mathcal{F}_1[\bm{\alpha}(t)]  
\end{eqnarray} 
A solution in a closed form, i.e., an exact solution can only be obtained by
taking the derivatives of $\mathcal{F}$ up to  $n^{\text{th}}$ order. This
approach leads to an infinite number of equations of motion for a hierarchical
set of auxiliary influence functionals $\mathcal{F}_n$.

The time derivatives of $\Lambda^{\times}[\alpha (\tau)]$ in general may yield
terms that are non-hierarchical and thus not all forms of the bath correlation
function may be employed to obtain hierarchically linked EOMs. The form of the
bath correlation functions such as the one based on the exponential decomposition
scheme or the Chebyshev decomposition scheme ensures the hierarchical form of
the derivatives of $\mathcal{F}$. The leading order of the system-bath coupling
of the auxiliary influence functional $\mathcal{F}_n$ is $2 n$. Thus memory effects
as well as the influence of higher orders of the system-bath coupling are
included in each tiers of the HEOM scheme.


In the present study, the Chebyshev decomposition scheme is utilized to find a closed solution to
the above problem by deriving  hierarchically linked equations of motion of the
auxiliary density matrices. As mentioned earlier, the starting point in the
Chebyshev decomposition is the utilization of the Chebyshev polynomials and
Bessel functions to express the bath correlation function in terms of a time
dependent and independent part: $C_j(t) = I_{j,n} C_{j,n}(t)$. Feynman
showed~\cite{feyn63a} that by employing stochastic bath variables that obey
Gaussian statistics, the bath averaged influence functional can be written in
terms of the bath correlation function, $C_j(t-\tau)$: 
\begin{eqnarray}
\mathcal{F}[\bm{\alpha}(t)] &=& \text{exp}\Big\{-\sum_{j} \int_{t_i}^{t} d\tau K_{j}^{\times}[\bm{\alpha}(\tau)]~ \Lambda_j^{\times}[\bm{\alpha}(\tau)] \Big\} ~. \label{f0}
\end{eqnarray}
%
where %
\begin{eqnarray}
\Lambda_j[\bm{\alpha}(\tau)] = \int_{t_0}^{t} d\tau C_j(t-\tau) K_j[\bm{\alpha}(\tau)]~.
\end{eqnarray}\\
%
Taking the time derivative of the Influence functional, Eq.~\ref{f0}, one gets %
\begin{eqnarray}
\partial_t \mathcal{F}[\bm{\alpha}(t)] &=& -\sum_{j} 
K_{j}^{\times}[\bm{\alpha}(t)]~  \Lambda_j^{\times}[\bm{\alpha}(t)]~
\mathcal{F}[\bm{\alpha}(t)]  ~. \label{f1}\\
&=& -\sum_{j} K_{j}^{\times}[\bm{\alpha}(t)]~
(\Lambda_{j}[\bm{\alpha}(t)]\mathcal{F}[\bm{\alpha}(t)]  -
\mathcal{F}[\bm{\alpha}(t)] \Lambda_{j}^{\dagger}[\bm{\alpha}(t)]) ~.\\\nonumber
%
&=&-\sum_{j}  K_{j}^{\times}[\bm{\alpha}(t)] \Big(\int_{t_0}^{t} d\tau C_j(t-\tau) K_j[\bm{\alpha}(t)] \mathcal{F}[\bm{\alpha}(t)]  - \mathcal{F}[\bm{\alpha}(t)] \int_{t_0}^{t} d\tau K_j[\bm{\alpha}(t)] C_j^{\dagger}(t-\tau) \Big) ~.\nonumber
\end{eqnarray} 
Finally, using the Chebyshev form of the bath correlation functions, i.e.,
$C_j(t) = \sum_k I_{j,k} C_{j,k}(t)$, results in
\begin{align}
\partial_t& \mathcal{F}[\bm{\alpha}(t)]= -\sum_{j}  K_{j}^{\times}[\bm{\alpha}(t)]\\\nonumber
&\times \Big(\int_{t_0}^{t} d\tau \sum_k I_{j,k} C_{j,k}(t-\tau) K_j[\bm{\alpha}(t)] \mathcal{F}[\bm{\alpha}(t)]  - \mathcal{F}[\bm{\alpha}(t)] \int_{t_0}^{t} d\tau K_j[\bm{\alpha}(t)] \sum_k I_{j,k}^{\dagger}C_{j,k}^{\dagger}(t-\tau) \Big) ~.
\end{align} 
At this point we introduce auxiliary influence functionals $\mathcal{F}_{j_1}[\bm{\alpha}(t)]$, such that $$\mathcal{F}_{j_1}[\bm{\alpha}(t)] = \Lambda_{j_1}[\bm{\alpha}(t)] \mathcal{F}[\bm{\alpha}(t)]~.$$
%
%
%
Using these auxiliaries, the time derivative of the influence functional can be rewritten as
\begin{eqnarray}
\partial_t \mathcal{F} [\bm{\alpha}(t)]= -\sum_{j} 
K_j^{\times}[\bm{\alpha}(t)]~ \big{[}~\mathcal{F}_{j_1}[\bm{\alpha}(t)]  -
\mathcal{F}[\bm{\alpha}(t)] \Lambda_j^{\dagger}[\bm{\alpha}(t)]~ \big{]}~,
\label{deriv_inf-fnc} 
\end{eqnarray} 
where
\begin{eqnarray}
\mathcal{F}_{j_1}[\bm{\alpha}(t)] = \int_{t_0}^{t} d\tau~ C_{j_1}(t-\tau)~\mathcal{F}[\bm{\alpha}(t)] K_{j_1}[\bm{\alpha}(t)]~.
\end{eqnarray} 
To be able to proceed, one needs to calculate the derivative of $\mathcal{F}_{j_1}(t)$ for
which the auxiliaries $\mathcal{F}_{j_1}(t)$ are now written also as decomposed
into time-independent and time-dependent parts
\begin{eqnarray}
\mathcal{F}_{j_1}[\bm{\alpha}(t)] = \sum_k I_{j,k} \mathcal{F}_{j_1,k}[\bm{\alpha}(t)] ~.
\end{eqnarray} 
Subsequently, one employs the time derivatives of $\mathcal{F}_{j_1,k}[\bm{\alpha}(t)]$ given by
\begin{eqnarray}
\partial_t \mathcal{F}_{j_1,k}[\bm{\alpha}(t)] &=&  \partial_t(\Lambda_{j_1,k}[\bm{\alpha}(t)] \mathcal{F}[\bm{\alpha}(t)] ) \\\nonumber
&=& \partial_t(\Lambda_{j_1,k}[\bm{\alpha}(t)])\mathcal{F}[\bm{\alpha}(t)] + \Lambda_{j_1,k}[\bm{\alpha}(t)] \partial_t(\mathcal{F}[\bm{\alpha}(t)] )  ~.  
\label{deriv_fn1}
\end{eqnarray} 
Thus, the derivative of $\mathcal{F}_{j_1}[\bm{\alpha}(t)]$ can be written as
\begin{eqnarray}
\label{f2}
\partial_t \mathcal{F}_{j_1}[\bm{\alpha}(t)]  &= &
\partial_t(\Lambda_{j_1}[\bm{\alpha}(t)]\mathcal{F}[\bm{\alpha}(t)] )
\\\nonumber & =& \partial_t(\sum_k I_{j_1,k} \Lambda_{j_1,k}[\bm{\alpha}(t)]
\mathcal{F}[\bm{\alpha}(t)]) = \sum_k I_{j_1,k}\partial_t(
\Lambda_{j_1,k}[\bm{\alpha}(t)] \mathcal{F}[\bm{\alpha}(t)]) ~.
\end{eqnarray} 
To proceed further,  one needs the derivatives of the time dependent quantities
in Eq. 30, i.e., $\Lambda_{j_1,k}[\bm{\alpha}(t)]$ and
$\mathcal{F}[\bm{\alpha}(t)]$. %
The time derivative of the term  $\Lambda_{j_1,k}[\bm{\alpha}(t)]$ is obtained
using the time derivatives of the respective Bessel functions. The time
derivative of $\mathcal{F}[\bm{\alpha}(t)]$ is given in Eq.~\ref{deriv_inf-fnc}.
Plugging it into Eq.~\ref{f2} above, one gets the following expression
\begin{align}
\partial_t& \mathcal{F}_{j_1,k}[\bm{\alpha}(t)] \\\nonumber= &
\partial_t(\Lambda_{j_1,k}[\bm{\alpha}(t)])\mathcal{F}[\bm{\alpha}(t)] +
\Lambda_{j_1,k}[\bm{\alpha}(t)] \big{(} -\sum_{j_2}
K^{\times}_{j_2}[\bm{\alpha}(t)] [\Lambda_{j_2}[\bm{\alpha}(t)]
\mathcal{F}[\bm{\alpha}(t)]  - \mathcal{F}[\bm{\alpha}(t)] 
\Lambda_{j_2}^{\dagger}[\bm{\alpha}(t)] ] \big{)}  ~.
\end{align} 
Moving $\Lambda_{j_1,k}[\bm{\alpha}(t)]$ inside the sum and identifying already defined operators in the resulting equations, one gets
\begin{align}
\partial_t \mathcal{F}_{j_1,k}[\bm{\alpha}(t)] &= (\partial_t
\Lambda_{j_1,k}[\bm{\alpha}(t)])\mathcal{F}[\bm{\alpha}(t)] \\\nonumber 
&-\sum_{j_2} K^{\times}_{j_2}[\bm{\alpha}(t)]~ [ \Lambda_{j_1,k}[\bm{\alpha}(t)]
\Lambda_{j_2}[\bm{\alpha}(t)] \mathcal{F}[\bm{\alpha}(t)] -
\underbrace{\Lambda_{j_1,k}[\bm{\alpha}(t)] 
	\mathcal{F}[\bm{\alpha}(t)]}_{\mathcal{F}_{j_1,k}[\bm{\alpha}(t)]}
\Lambda_{j_2}^{\dagger}  ]   ~.\\\nonumber
&=  (\partial_t \Lambda_{j_1,k}[\bm{\alpha}(t)])\mathcal{F}[\bm{\alpha}(t)] \\\nonumber 
&-\sum_{j_2} K^{\times}_{j_2}[\bm{\alpha}(t)]~ [\sum_{k'} I_{j_2,k'}
\underbrace{\Lambda_{j_1,k}[\bm{\alpha}(t)] \Lambda_{j_2,k}[\bm{\alpha}(t)]
	\mathcal{F}[\bm{\alpha}(t)]}_{\mathcal{F}_{j_1,j_2,k,k'}[\bm{\alpha}(t)]} -
\underbrace{\Lambda_{j_1,k}[\bm{\alpha}(t)] 
	\mathcal{F}[\bm{\alpha}(t)]}_{\mathcal{F}_{j_1,k}[\bm{\alpha}(t)]}
\Lambda_{j_2}^{\dagger}  ]   ~.\\\nonumber
&=  (\partial_t \Lambda_{j_1,k}[\bm{\alpha}(t)])\mathcal{F}[\bm{\alpha}(t)] \\\nonumber
& -\sum_{j_2} K^{\times}_{j_2}[\bm{\alpha}(t)]~ [ \underbrace{ \sum_{k'} I_{j_2,k'} \mathcal{F}_{j_1,j_2,k,k'}[\bm{\alpha}(t)]}_{\mathcal{F}_{j_1,j_2}[\bm{\alpha}(t)]} - \underbrace{\Lambda_{j_1,k}[\bm{\alpha}(t)]  \mathcal{F}[\bm{\alpha}(t)]}_{\mathcal{F}_{j_1,k}[\bm{\alpha}(t)]} \Lambda_{j_2}^{\dagger}  ]   ~.\nonumber
\end{align} 
The final form of the first-tier ADM is therefore given by
\begin{align}
\partial_t \mathcal{F}_{j_1,k}[\bm{\alpha}(t)] &=  (\partial_t \Lambda_{j_1,k}[\bm{\alpha}(t)])\mathcal{F}[\bm{\alpha}(t)] \\\nonumber
& -\sum_{j_2} K^{\times}_{j_2}[\bm{\alpha}(t)]~ [\sum_{k'} I_{j_2,k'} \mathcal{F}_{j_1,j_2,k,k'}[\bm{\alpha}(t)] - \mathcal{F}_{j_1,k}[\bm{\alpha}(t)] \Lambda_{j_2}^{\dagger}[\bm{\alpha}(t)]  ]   ~.\nonumber
\end{align}
Time derivatives of the second-tier AIF which can be found in the above derivative of the first-tier AIF and are defined is as follows
\begin{align}
\partial_t \mathcal{F}_{j_1,j_2,k,k'}[\bm{\alpha}(t)] &=  (\partial_t\Lambda_{j_1,k}[\bm{\alpha}(t)])\Lambda_{j_2,k'}[\bm{\alpha}(t)] \mathcal{F}[\bm{\alpha}(t)]\\\nonumber
& + \Lambda_{j_1,k} [\bm{\alpha}(t)] ~(\partial_t \Lambda_{j_2,k'}[\bm{\alpha}(t)])\mathcal{F}[\bm{\alpha}(t)]\\\nonumber
& -\sum_{j3} K^{\times}_{j3}~ [\sum_{k''} I_{j3,k''} \mathcal{F}_{j_1,j_2,j3,k,k',k''}[\bm{\alpha}(t)] - \mathcal{F}_{j_1,j_2,k,k'}[\bm{\alpha}(t)] \Lambda_{j3}^{\dagger}  ]   ~.\nonumber
\end{align} 
Using the simplified notation $\mathcal{F}[\alpha(t)]$ = $\mathcal{F}(t)$ and
noting that the $(n)$ in the superscript denotes the level of the hierarchy
while the $k$ in the subscript denotes the Chebyshev polynomials, the general
form of the AIF is  given by
\begin{eqnarray}
\mathcal{F}_{j_m, k}^{(n)}(t)= \prod_{j_m,k} ( \Lambda_{j_m,k}(t) ) \mathcal{F}(t)~.
\end{eqnarray}
Here $m$ denotes the site index, and $\Lambda_{j_m}(t)$ is given as
\begin{eqnarray}
\Lambda_{j_m}(t) = \sum_n I_{j_m,k}~ \Lambda_{j_m,k}(t) ~.
\end{eqnarray} 
The time derivative of the general form, or the $n^{th}$ tier of the AIF, can be
calculated to yield
\begin{align}
\partial_t \mathcal{F}_{j_{m},k}^{(n)} = & - \sum_{j}  K_{j}^{\times} [\bm{\alpha}(t)] \big( \sum_{k}^N I_{j_m,k} \mathcal{F}_{j_{m},k}^{(n+)} - \mathcal{F}_{j_{m},k}^{(n)} \Lambda_{j_m}^{\dagger}[\bm{\alpha}(t)] \big)\\\nonumber
& - i\sum_{r=1}^n \big( \bar{\omega}\mathcal{F}_{j_{m},k}^{(n)} \big)\\\nonumber
&+ \begin{cases}
 \Omega \sum_{r=1}^{n} \big( - \mathcal{F}_{j_{m},k}^{(n)}\big)+ \sum_{r=1}^{n}  \big(K_j \mathcal{F}_{j_{m},k}^{(n-)} \big)~~~~~~~~~~~~~~~\text{if}~ k = 0\\
  \frac{\Omega}{2}\sum_{r=1}^{n} \big(\mathcal{F}_{j_{m},k-1}^{(n)} -  \mathcal{F}_{j_{m},k+1}^{(n)} \big)~~~~~~~~~~~~~~~~~~~~~~~~~~~~\text{if}~0< k < K_{ch}\\
 \frac{\Omega}{2}\sum_{r=1}^{n} \big( \mathcal{F}_{j_{m},k-1}^{(n)}\big)~~~~~~~~~~~~~~~~~~~~~~~~~~~~~~~~~~~~~~~~~~\text{if}~n = K_{ch}
\end{cases}~.
\end{align}
The obtained infinite hierarchy of influence functionals leads to an infinite
hierarchy of auxiliary reduced density matrices
\begin{align}
\bm{\rho}^{(n)}(t) = \mathcal{U}_n(\alpha, \alpha', t; {\alpha_0, \alpha'_0, t_0}) \bm{\rho}^{(n)}(t_0)~,
\end{align}
where the auxiliary time evolution operator $\mathcal{U}_n$ is obtained by
replacing the influence functional $\mathcal{F}$ in Eq.~\ref{time_evl_op} by the corresponding
auxiliary influence functional $\mathcal{F}_n$. The time evolution of the RDM
and the ADMs can thus be expressed in terms of a system of coupled differential
equations
\begin{align}
\partial_t \bm{\rho}_{j_{m},k}^{(n)} =& -i\mathcal{L}_{S}(t) \bm{\rho}_{j_{m},k}^{(n)}\\\nonumber
& - \sum_{j}  K_{j}^{\times} \big( \sum_{n}^N I_{j_m,k} \bm{\rho}_{j_{m},k}^{(n+)} - \bm{\rho}_{j_{m},k}^{(n)} \Lambda_{j_m}^{\dagger} \big)\\\nonumber
& - i\sum_{r=1}^n \big( \bar{\omega}\bm{\rho}_{j_{m},k}^{(n)} \big)\\\nonumber
&+ \begin{cases}
 \Omega \sum_{r=1}^{n} \big( - \bm{\rho}_{j_{m},k}^{(n)}\big)+ \sum_{r=1}^{n}  \big(K_j \bm{\rho}_{j_{m},k}^{(n-)} \big)~~~~~~~~~~~~~~~\text{if}~ n = 0\\
  \frac{\Omega}{2}\sum_{r=1}^{n} \big( \bm{\rho}_{j_{m},k-1}^{(n)} -  \bm{\rho}_{j_{m},k+1}^{(n)} \big)~~~~~~~~~~~~~~~~~~~~~~~~~~~~\text{if}~0< k < K_{ch}\\
 \frac{\Omega}{2}\sum_{r=1}^{n} \big( \bm{\rho}_{j_{m},k-1}^{(n)}\big)~~~~~~~~~~~~~~~~~~~~~~~~~~~~~~~~~~~~~~~~~~\text{if}~k = K_{ch}
\end{cases}~.
\end{align}
where $\mathcal{L}_{S} \bullet = [H_S, \bullet]$ denotes the Liouville
superoperator.

%
Since the above HEOM has potentially infinite orders, it needs to be truncated
at a certain level of hierarchy. Different truncation
schemes\cite{klei04a,xu05a,schr06c,chen09b} exist that cut the HEOM at a
specific level of hierarchy with the resulting equations corresponding to
different orders of perturbative QMEs. Truncating the HEOM equations at a given
level $n$ results in the perturbation order $N=2n$. If the ADOs whose level is
higher than $n$ are simply discarded, one obtains the so-called time-nonlocal
truncation. For $n=1$ one gets the second-order TNL
QME \cite{schr06c}. This truncation scheme has been shown for several test cases to lead to spurious oscillations in the
population dynamics. Another scheme developed by Xu and Yan yields the time-local approximation \cite{xu05a}  for the leading tier of the HEOM. With regard to
truncating the Chebyshev HEOM, this amounts to approximating the term of the
leading tier using the following equation
\begin{eqnarray}
\sum_{k} \rho^{(n+)}_{j_{m},k} = -i [\Lambda_{j_m} \rho^{(n)}_{j_{m},k}]
\end{eqnarray}
where
\begin{eqnarray}
\Lambda_{j_m} = \int_{t_0}^{t} d\tau C_{j_m} (t-\tau) e^{-i H_S \tau} K_j e^{i H_S \tau}~.
\end{eqnarray}
Below we will denote the first-tier  HEOM which is second order in the
system-bath coupling as TL2 and TNL2 for the time-local and the time-nonlocal
truncation schemes, respectively.   The second-tier variants which are fourth
order in system-bath coupling will be denoted as TL4 and TNL4. In an earlier
study \cite{pope16a}, the C-HEOM TL2 has also been termed CDTL, i.e., Chebyshev
decomposition time-local (second-order in system-bath coupling).

\section{Results and Discussion} \label{results_heom} %
In the following we apply the Chebyshev-HEOM to calculate the dynamics of a
dissipative two-level model system. This arrangement has been extensively
studied as a benchmark system \cite{ishi10a,ishi11a, aght12a}. At higher
temperatures, the dephasing in this modeland similar models can even reasonably
be described by classical noise  \cite{aght12a,hu18a}. In the following, we show
and discuss results for the population dynamics obtained using the first and the
second tier of the time-non-local (TNL2 and TNL4) as well as the time-local
truncation (TL2 and TL4) of the C-HEOM together with  converged results
calculated using the exponentially decomposed HEOM as well as the
ML-MCTDH~\cite{wang03, wang08b} schemes. Some calculations using the Chebyshev
decomposition in a 2nd-order scheme that have been presented in a previous study
\cite{pope16a} are also included. In the model considered, each of the two
sites, with site energies $E_1$ and $E_2$ and coupling elements $V_{12} =
V_{21}$, is coupled to its own independent phonon bath, each of which is assumed
to have the same spectral density. While the C-HEOM can handle any kind of an
energy distribution or spectral density of the bath, as a first test case we
employ a Drude-Lorentz form of the system-bath coupling so that the scheme can
be tested against the exponential HEOM. Thereafter, we employ an Ohmic spectral
density and compare C-HEOM results with the ML-MCTDH scheme, followed by a
super-Ohmic spectral density that we employ to show the capabilities of the
C-HEOM. \subsection{Drude-Lorentz Spectral Density} The Drude-Lorentz form of
the spectral density is given by
\begin{equation}\nonumber
\mathcal{J}(\omega) = 2\pi\lambda \gamma \frac{\omega}{\omega^2 + \gamma^2}~,
\end{equation}
with the real parameters $\gamma$ and $\lambda$. Since each bath is connected to
its own thermal bath, the spectral densities could vary but for simplicity we
assume that the system-bath coupling is the same for all sites. The parameter
$\gamma$ denotes the inverse correlation time and determines the width of the
function, while the parameter $\lambda$ denotes the reorganization energy and
determines the amplitude of the function. For the simulations shown in this
section the correlation time is chosen to be $1/\gamma = 100$~fs, the
electronic coupling $V_{12}=V_{21}=100$~cm$^{-1}$ while the temperature $T$ is
set to 300~K. Two different configurations of the site energies are chosen for
the test below. Fig.~1 shows the case where the the site energies are equal,
i.e., $E_1 = E_2$, while Fig.~2 shows a biased case where the energy of the one
site is higher than that of the other one, i.e., $E_1 - E_2 = 100$~cm$^{-1}$.
For the above scenarios we show results for four different values of the
reorganization energy, which represents the strength of the electronic coupling,
ranging from low to high values. Each of the panels for the different values of
$\lambda$ shows the population dynamics of the first site given by the first
element $(1,1)$ of the density matrix. The converged exponential HEOM results
mentioned earlier are obtained from Ref.~\citenum{aght12a} while the C-HEOM TL2
results have been discussed already in Ref.~\citenum{pope16a}. The population
dynamics that are converged in the system-bath coupling are obtained by
increasing the number of tiers utilized in the HEOM scheme until no change in
the dynamics is visible anymore. The larger the reorganization energy, the 
higher the number of tiers in the HEOM scheme required to achieve convergence.

For the case of equal site energies, i.e.,  $E_1 = E_2$, one can see that for
the lowest reorganization energy, the agreement between all  approaches and
tiers is good. Since  the influence of the bath on the system is very low, even
the second-order perturbative treatment of the system-bath coupling can account
for the effects of the environment correctly. For $\lambda = 20$ cm$^{-1}$ the
level of damping in the oscillations shown by C-HEOM TL2, TNL2 and TNL4 starts
to differ from the converged results that are most closely obtained by TL4
C-HEOM. For the reorganization energy value set to 100~cm$^{-1}$, the TL4 C-HEOM
deviates from the converged results and remains close to the TL2 C-HEOM, while
the TNL4 and TNL2 C-HEOM results shows higher oscillations which are a
characteristic of the TNL approach, where the next tier of HEOM is simply put to
zero. Thus, the second-order perturbation theory, i.e., TL2 and TNL2, as well as
the fourth-order TNL4 and TL4 C-HEOM find their theoretical limits already in
the  intermediate coupling regime. For the highest coupling strength of
500~cm$^{-1}$, the TNL4 as well as the TL4 versions of the C-HEOM seem to decay
to erroneous values with the TNL version showing higher oscillations. As
compared to the converged results, TL2 C-HEOM appears to be overdamped and
decays to the thermal equilibrium quickly, while TNL2 C-HEOM predicts a dynamics
that decays too slow.

\begin{figure}[!t]
\centering
\includegraphics[width=0.9\textwidth]{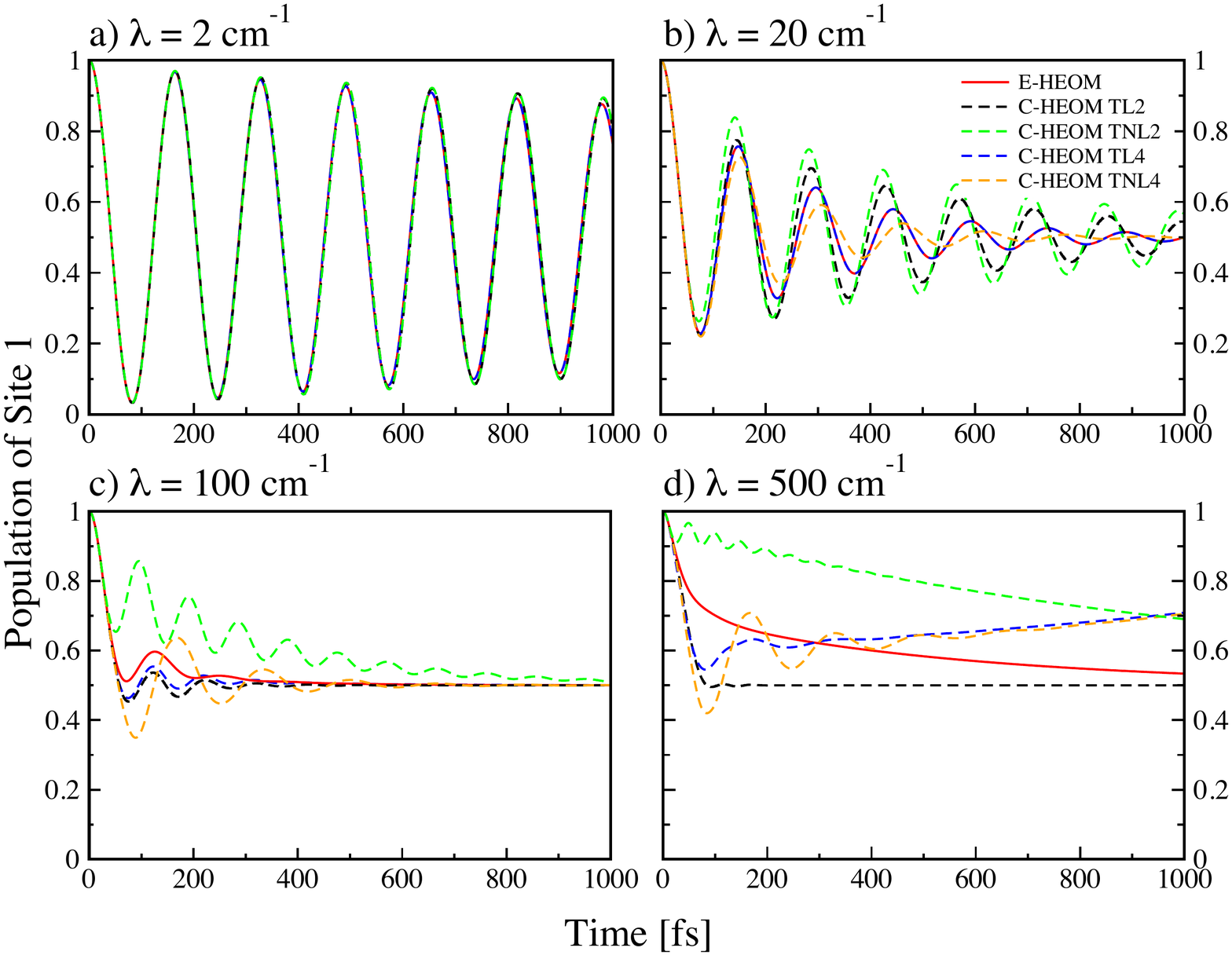} \caption[]{Time
evolution of the population on the first site which is initially populated for 
different reorganization energies. The site energies are set to equal values for this case.  \label{eq_site_en_figure}}
\end{figure}

For the case of unequal site energies shown in Fig. 2, the overall picture is
somewhat different. Similar to the above case, the fourth and even the
second-order perturbative approaches yield an accurate description of the level
of damping when the reorganization energy is low, i.e., $\lambda = 2$ cm$^{-1}$.
Already for intermediate coupling strengths, the TL4 C-HEOM can be seen to
deviate from the equilibrium value of the converged results although it follows
the dynamics of the converged E-HEOM closely for the initial part, at least for
$\lambda$ = 20~cm$^{-1}$. At the same time, the TNL4 C-HEOM results reach the
equilibrium value of the converged results more closely, albeit showing a lower
damping or higher oscillations. The second-order TL scheme is overdamped as
compared to converged E-HEOM values, while the TNL version predicts higher oscillation amplitudes. For the highest reorganization energy shown ($\lambda =
500$ cm$^{-1}$), the TL4 as well as TNL4 C-HEOM deviate strongly from the
equilibrium populations at longer times, while the TL2 and TNL2 C-HEOM seem to
decay to the correct equilibrium values. This indicates that using more tiers
of the HEOM does not necessarily provide better results, i.e., results closer to
the converged dynamics. As convergence is not achieved by neither the first nor
the second-tier approaches, the accuracy of the results is difficult to assess.
Before convergence is achieved, different tiers could behave differently, a
higher one not necessarily being better than the lower ones, and each may show
its own peculiar behavior. Successive tiers of the HEOM may show convergence to
a particular converged value for a chosen, specific set of parameters, but might behave completely different in another parameter regime. %
\begin{figure}[!t]  
\includegraphics[width=0.9\textwidth]{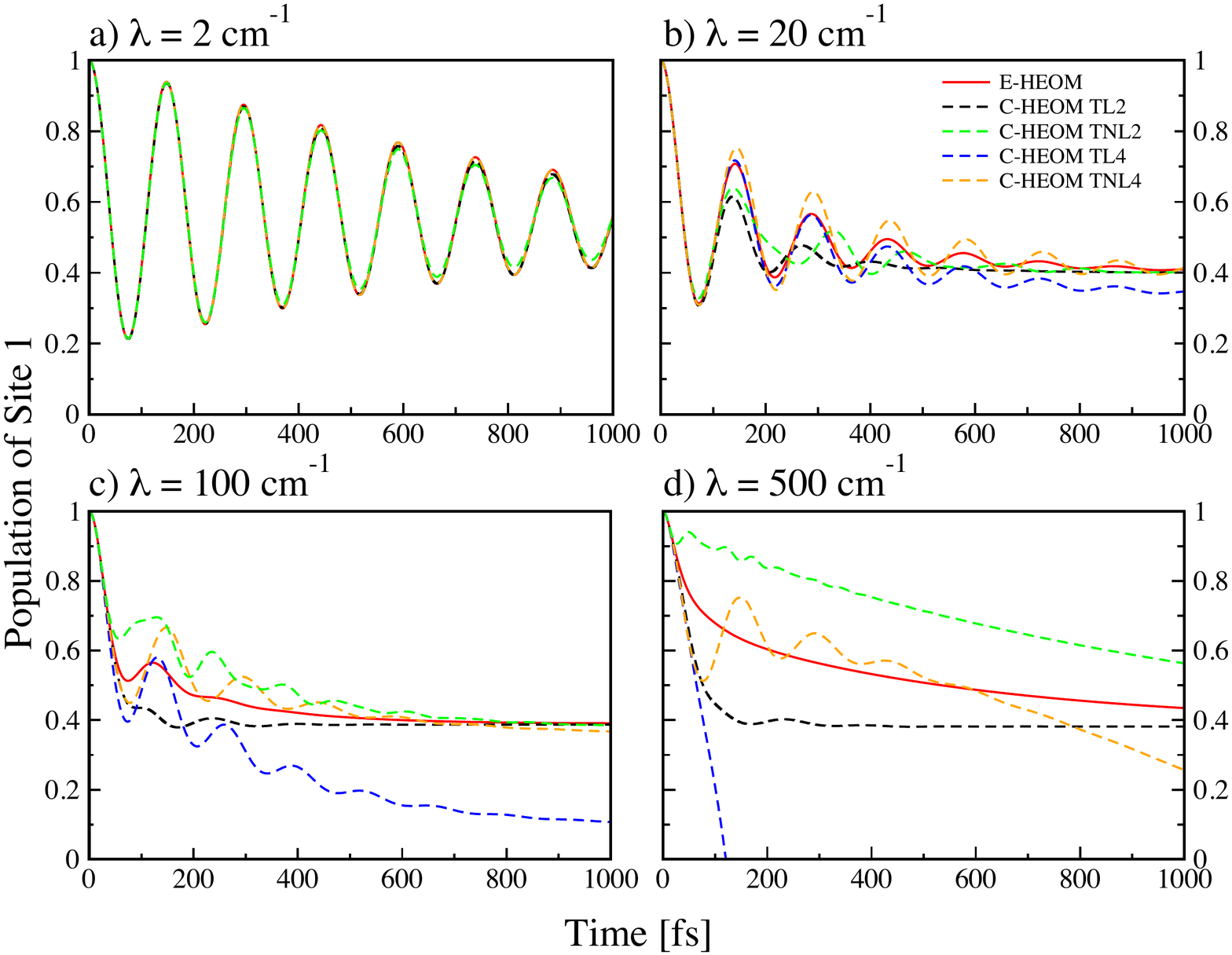} \caption[]{Time
evolution of the population on the first site which is initially populated for 
different reorganization energies. The site energies are in a biased configuration,
i.e., $\Delta E = E_2 - E_1 = 100$ cm$^{-1}$. \label{eqtwo_site_en_figure}}
\end{figure}
Overall one can see that the TL4 or TNL4 C-HEOM, which represent the fourth
order in perturbation theory, do  perform better only for a small range of
system-bath coupling strength.  For low to intermediate coupling strengths, the
second tier or fourth order can be seen to perform better than the second order
schemes, but for higher values of the reorganization energy one needs an
enlarged number of tiers of the HEOM in order to achieve convergence. Moreover,
the  TNL2 as well as TNL4 results are obtained both from the Chebyshev version
of the HEOM as well as the exponential scheme and are equal (data not shown).
Thus, we can conclude that the presented C-HEOM approach is a valid scheme and
provides the same results as the E-HEOM for scenarios in which both of them can
be applied without further approximations. In this work we conducted simulations
only for the second tier truncation of the C-HEOM, which can in principle be
extended to higher tiers. Memory constraints would not allow the extension of
the scheme to a high number of tiers and thus the formalism may be inapplicable
to strong coupling regimes unless the computational load is somehow reduced by,
for example, implementing the C-HEOM scheme on parallel processing units. %
%
\subsection{Ohmic Spectral Density}

Next we apply the time-local versions TL2 and TL4 as well as the time non-local
version TNL2 of the C-HEOM approach to the dissipative dynamics in two level
systems at zero temperature. In this case, the system  is in contact with a
bosonic bath that is described by an Ohmic  spectral density  given by
\begin{eqnarray}
\mathcal{J}(\omega) = \frac{\pi \lambda \omega}{4 \omega_c} \text{exp} \Bigg( {\frac{-\omega}{\omega_c}} \Bigg)~.
\end{eqnarray}
As a benchmark we utilize  the ML-MCTDH approach \cite{wang03, wang08b} to
provide converged results of quantum dissipative dynamics as reported in
Ref.~\cite{tang15a}. This test is carried out in order to further demonstrate
the capability and reliability of the C-HEOM formalism. Specifically, it should
be shown that the numerical cost of the C-HEOM remains constant at all
temperatures, including $T$ = 0~K, a property that sets the C-HEOM completely
apart from the exponential HEOM. With decreasing temperature the exponential
HEOM requires more terms from the Matsubara or Pade expansion to achieve
convergence and to accurately obtain the dynamics. Thus, at low temperatures the
application of the exponential HEOM is severely limited. While the extended
HEOM~\cite{tang15a} is expected to be available at an arbitrary temperature
under the condition that a proper expansion set is applied, the C-HEOM is
available at all temperatures without any accompanying restrictions.

As in the previous section, two different configurations of the site energies
have been chosen. Figure~3 shows the equal site energy case, i.e., $E_1 = E=2$,
while Fig.~4 shows the biased case where the energy of the first site is higher,
i.e., $E_1 - E_2 = 100$ cm$^{-1}$. The initial population of site one is again
set to one. The calculations have been performed using the C-HEOM approach for
different sets of parameters $\Delta$, i.e., the coupling strength between the
sites, and $\lambda$, the reorganization energy. The same parameters sets as in
Ref.~\citenum{tang15a} were selected to be able to compare to the converged
ML-MCTDH results.


For the cases with the largest inter-site coupling, i.e.,   $\Delta$ =
100cm$^{-1}$,  the C-HEOM results improve with the  order of the hierarchy and
gradually converge to the benchmark result obtained by the ML-MCTDH approach.
These cases are depicted in panels c) and d) of Figs.~4 and  5. For $\Delta E$ =
$E_1$ - $E_2$ = 0, $\Delta$ = 40 cm$^{-1}$, and $\lambda$ = 40 cm$^{-1}$ shown
in Fig. 4~a, the TL4 findings seem to follow the converged results the best,
albeit being slightly overdamped. The TL2 findings, while retaining the correct
damping strength, deviate slightly in phase. If one cranks up the reorganization
energy $\lambda$, as shown in Fig. 4~b, the TL4  results follow the converged
results much better than the TL2 ones. For the biased site energy case shown in
Fig. 5, with $\Delta$ = 20cm$^{-1}$, and $\lambda$ = 40cm$^{-1}$, the TNL2
scheme behaves most accurately, while for higher reorganization energy $\lambda$
the TL2 findings mimic the converged results closely. As already discussed
above, improving the tier by one order does not always directly increase the
results as can be seen for this case in the transition from TL2 to TL4.


\begin{figure}[!t]
\centering \includegraphics[width=0.9\textwidth]{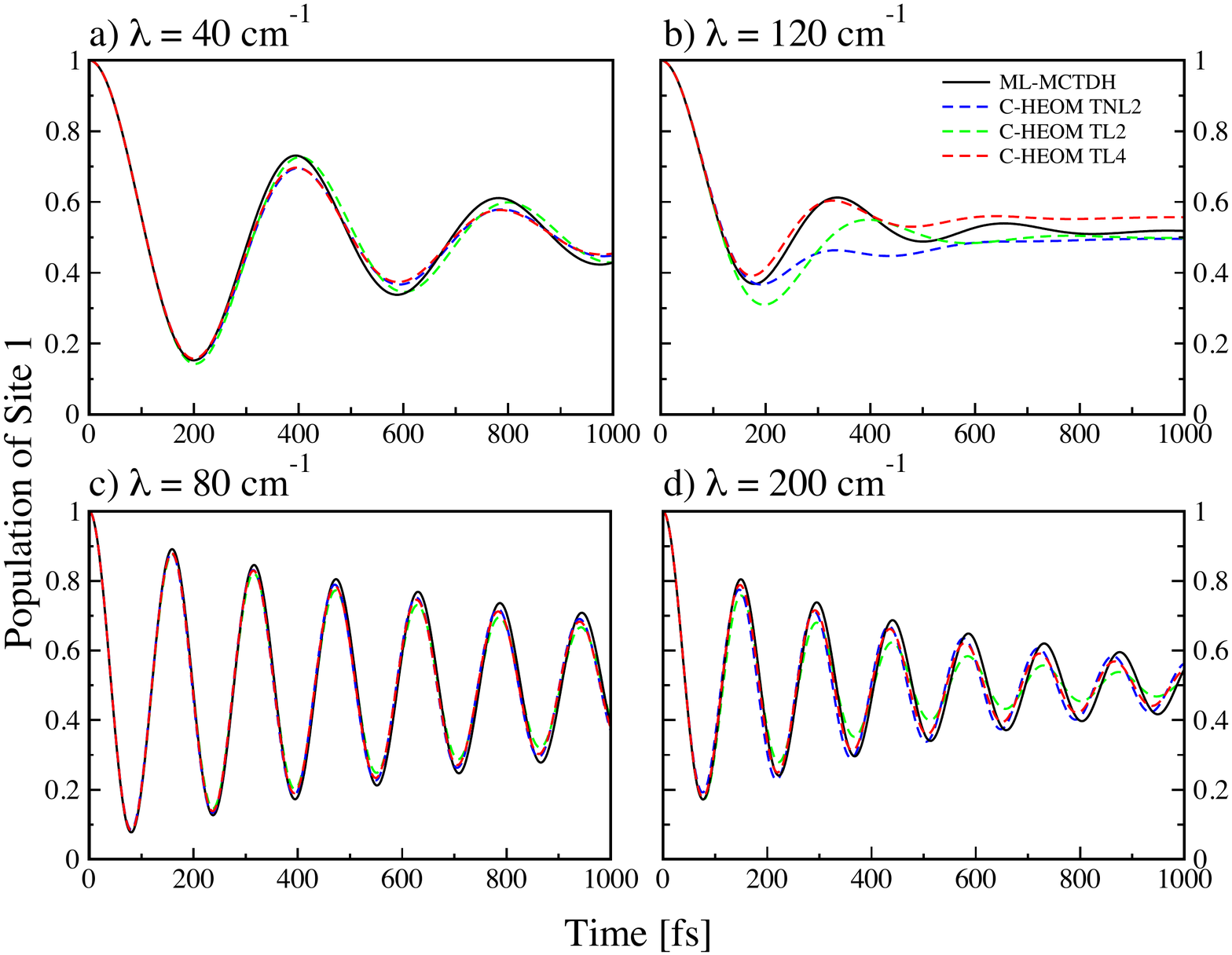}
\caption[]{Population dynamics of the initially populate site.  Quantum Ohmic
	noise ($\omega_c$ = 10 ps$^{-1}$ $\approx$ 53 cm$^{-1}$) is employed at zero temperature: (a) $\lambda$
	= 40 cm$^{-1}$ and (b) $\lambda$ = 120 cm$^{-1}$ (both with $\Delta$ = 40
	cm$^{-1}$); (c) $\lambda$ = 80 cm$^{-1}$ and (d) $\lambda$ = 200 cm$^{-1}$ (both
	with $\Delta$ = 100 cm$^{-1}$). The site energies are set to equal values.
	\label{ohm_eq_zero_figure}}
\end{figure}

\begin{figure}[!t]
\centering \includegraphics[width=0.9\textwidth]{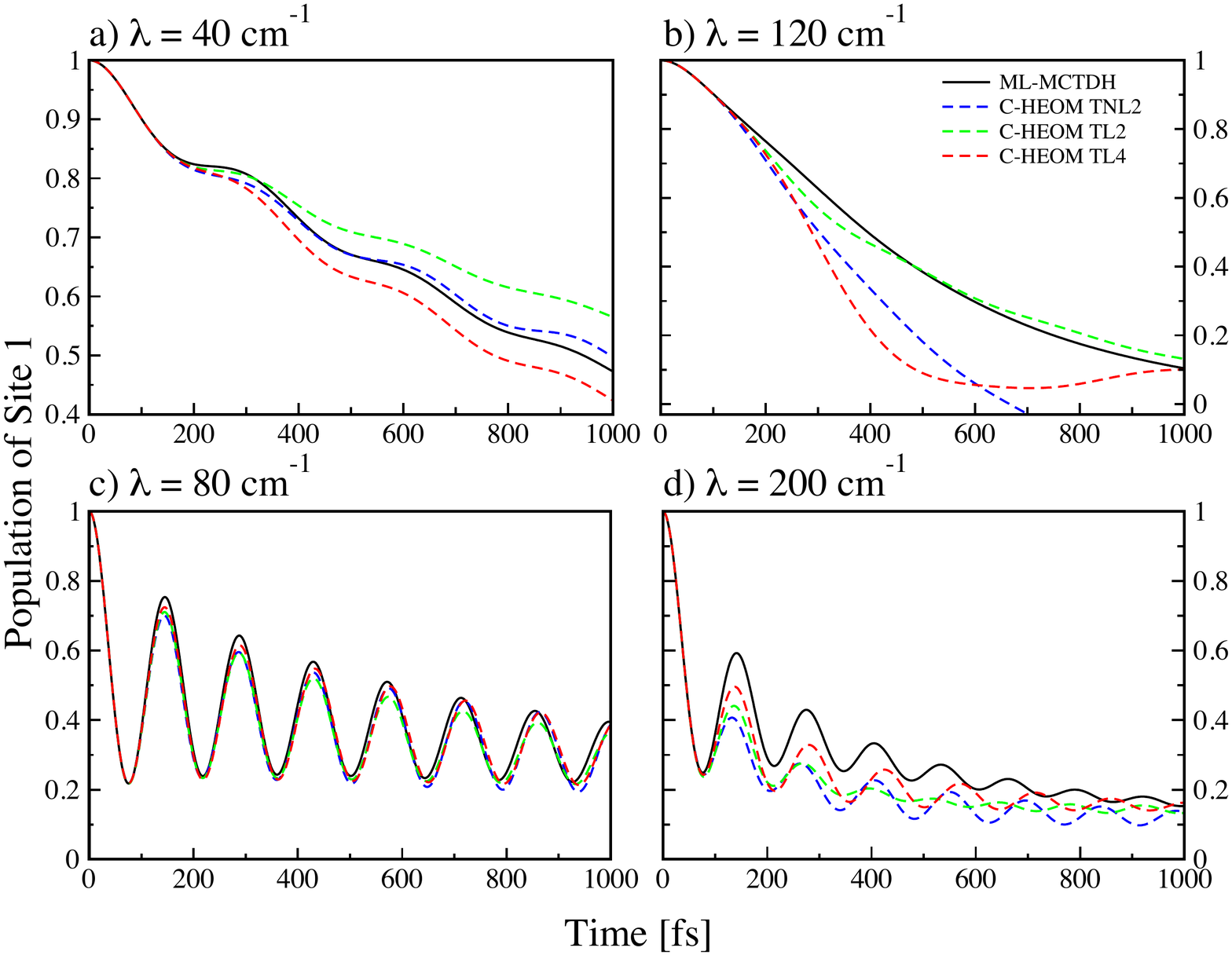}
\caption[]{Population dynamics of the  site which is initially
	populated in a quantum Ohmic noise environment ($\omega_c$ = 10 ps$^{-1}$ $\approx$ 53 cm$^{-1}$) at zero
	temperature: (a) $\lambda$ = 40 cm$^{-1}$ and (b) $\lambda$ = 120~cm$^{-1}$
	(both with $\Delta$ = 20 cm$^{-1}$) ; (c) $\lambda$ = 80 cm$^{-1}$ and (d)
	$\lambda$ = 200 cm$^{-1}$ (both with $\Delta$ = 100 cm$^{-1}$). A bias
	was applied to the site energies, i.e., $\Delta E$ = $E_1$ - $E_2$ =
	100~cm$^{-1}$~.  \label{ohm_uneq_zero_figure}}
\end{figure}

\subsection{Super-Ohmic Spectral Density}

To obtain a spectral density with a super-Ohmic character, we utilize the
following function, setting the value of $s$ to 4
\begin{eqnarray}
\mathcal{J}(\omega) = \frac{\pi \lambda s}{4 \Gamma(1+s)} \Bigg{(}\frac{\omega}{\omega_c} \Bigg{)}^S \text{exp}  \Bigg( \frac{-\omega}{\omega_c}  \Bigg)~.
\end{eqnarray}
According to the value of $s$, the spectral density given above can be classified into three categories
\begin{itemize}
\item 0 $<$ $s$ $<$ 1 : sub-Ohmic;
\item $s$ = 1 : Ohmic;
\item $s$ $>$ 1 : super-Ohmic:
\end{itemize}
While the Ohmic spectral density ($s$ = 1) has been thoroughly investigated by
analytical and numerical approaches, the dynamics for the sub-Ohmic (0$ <$ $s$
$<$ 1) and especially for the super-Ohmic ($s$ $>$ 1) cases are less thoroughly
understood \cite{wang16b,sun16b}.
 Using the C-HEOM, the super-Ohmic case can be
studied at the numerical cost of other simpler spectral densities, even at zero
temperature. In  Fig.~5 we show results using the TL2 and TL4 schemes  as well
as  TNL2 using C-HEOM for a two-site system at zero temperature attached to a
super-Ohmic environment. To this end, we study the above system for two
different values of reorganization energy $\lambda$. The parameter $\omega_c$ is
kept at 10~ps$^{-1}$, while the electronic coupling is set at
$V_{12}=V_{21}=100$~cm$^{-1}$. For $\lambda$ = 2~cm$^{-1}$ one obtains no
difference between the aforementioned tiers tested here, thus for low
reorganization energies one obtains converged results using even the second
order in perturbation theory. This is shown in Fig.~5 by the curves showing low
damping. Moreover,we tested the initial tiers of the C-HEOM for a reorganization
energy $\lambda$ = 30~cm$^{-1}$ shown in Fig.~5 by the lines that show a higher
damping. As can be seen, TL2  slightly deviates from the TL4. As evident from
the previous section, for larger $\omega_c$, i.e., 100~cm$^{-1}$, the second
tier of the C-HEOM has more or less converged to the accurate results of the
ML-MCTDH approach for $\lambda$ = 80cm$^{-1}$). It is  therefore  reasonable to
assume that the results here for $\lambda$ = 30cm$^{-1}$ are not far from
convergence.

\begin{figure}[!t]
\centering \includegraphics[width=0.9\textwidth]{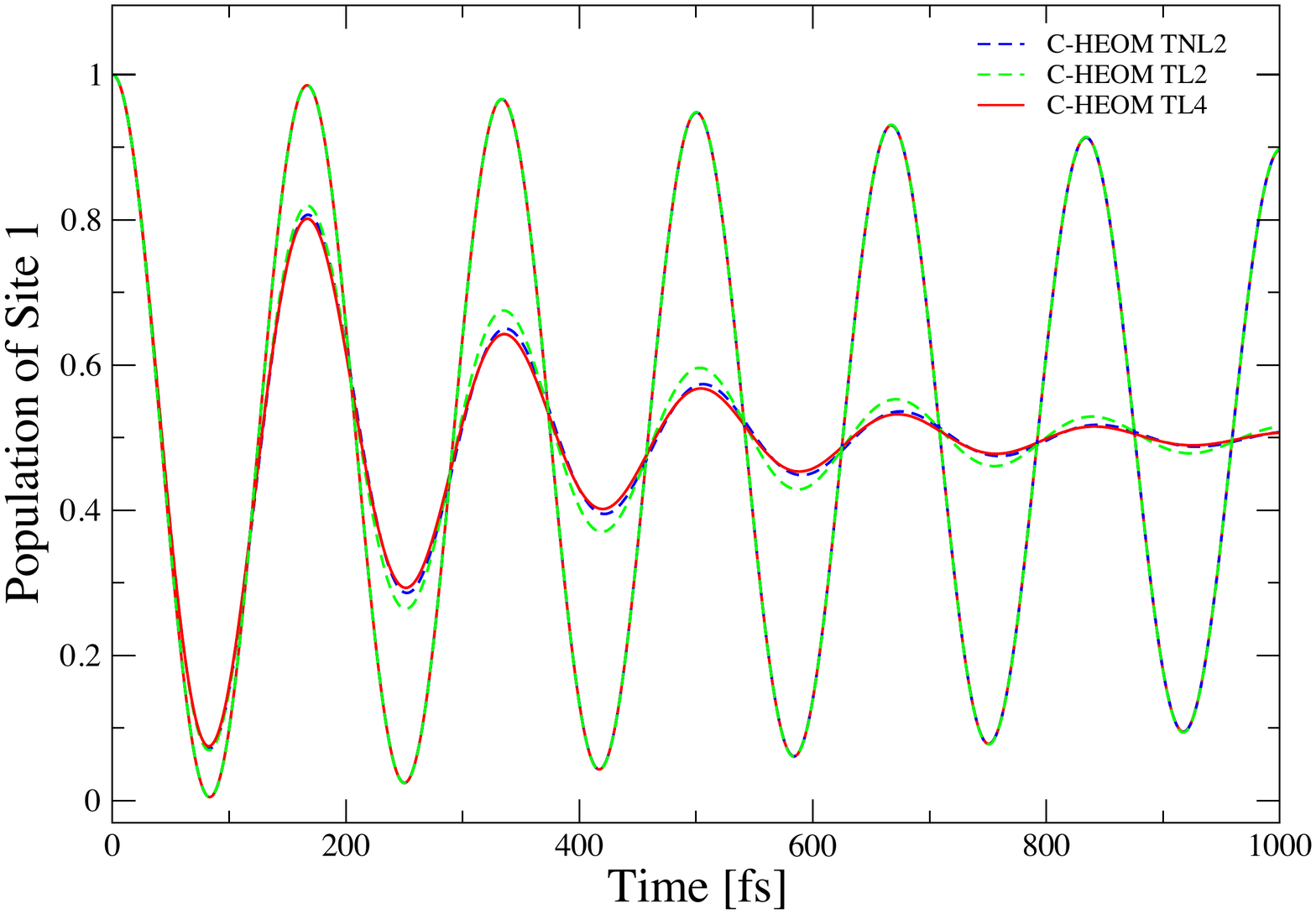}
\caption[]{Population dynamics of the site which is initially populated for a
	super-Ohmic spectral density at zero temperature for two different values of
	$\lambda$. The site energies are assumed the same. \label{super_ohm_figure}}
\end{figure}



\section{Conclusion}

In this paper we have introduced a new hierarchical equation of motion by
employing the Chebyshev decomposition of the bath correlation functions. The
C-HEOM approach presented forms an extension of the already reported work for
the second-order QME \cite{pope16a} which is identical to the first tier C-HEOM.
Here we have derived the general formula for calculating the C-HEOM to any
number of tiers  and presented numerical results for the first and second tier truncation
of the C-HEOM. We note that the simulation time using the C-HEOM scales linearly
with the number of Chebyshev polynomials utilized in the expansion of the bath
correlation functions. With regard to the second-order QME, this restriction
does not create any numerical hindrance since the sizes of the involved matrices
are small. In fact, for moderate simulation time the Chebyshev based QME, i.e.,
TL2 C-HEOM, approach in general takes up to two orders of magnitude less time to
generate the results compared to the exponential decomposed QME. However, the
story changes when one goes a higher number of tiers (or orders if one is
talking in terms of perturbation theory) in the EOMs. %
Already the second tier of the C-HEOM performs on a similar level as compared to
the E-HEOM as far as the time taken to simulate a certain simulation time is
concerned. In general for the Chebyshev based scheme, the higher the tier of the hierarchy, the larger the size of the auxiliary matrices it is composed of. For example, while the size of the
first tier auxiliary operators depends linearly on the number of Chebyshev
polynomials utilized, the size of the second-tier auxiliary operators scales
with the square of the number of Chebyshev polynomials, i.e., one has a
quadratic scaling. The situation worsens for three tiers, where the above
mentioned scaling becomes cubic. Thus, in general, if $n$ represents the tier of
the C-HEOM, and in fact also of the E-HEOM, then the size of auxiliary operator
scales as $\mathcal{O}^n$. In the case of the E-HEOM, where the scaling behavior
is the same, one does not see any practical hindrance as the simulated time does
not depend on the number of poles utilized in the exponential decomposition
scheme. The number of poles over which a sum is taken to reproduce the
correlation functions in terms of a sum of weighted exponential functions, the
number which is truncated at a certain point where one can observe converged
results, is affected only by the shape of the Bose-Einstein or the Fermi function or the temperature
of the bath(s), and the spectral density, i.e., the form or shape of the energy
distribution of the particles in the bath. Thus, for a fixed number of poles one
can obtain converged results for an arbitrary length of time. The case of the
Chebyshev decomposition scheme is, in a certain way, opposite to the case of the
exponential decomposition scheme. There is no effect of the shape of the energy distribution
function or the spectral density of the baths on the number of Chebyshev
polynomials needed to produce converged results. But while one gains the ability
to simulate arbitrary temperatures and forms of the bath density of states, one
can only obtain converged results up to a certain simulation time after
which the scheme breaks down. An estimate for terms needed to obtain converged
results can be given by \cite{coif93a,land11a,pope15a}
\begin{equation}
N = \Omega T + 10 \ln(\Omega T)~.
\end{equation}
The C-HEOM scheme has a bottleneck in the above mentioned scaling and is thus
only suitable for shorter to moderate lengths of time if one needs to employ
higher tiers of the C-HEOM. %
Nevertheless the crucial advantage of the C-HEOM, as mentioned earlier, over
approaches that utilize exponential functions to represent the correlation
functions, such as the original HEOM, is that there is no limitation on the form
of the spectral density or on the system temperature. Sharply fluctuating, even
step-like energy profiles of the environment at zero temperature can be dealt
with using the C-HEOM at no additional numerical cost as compared to single
Lorentzian forms at high temperatures. The advantage of the C-HEOM then is
twofold. Firstly, that the scheme can directly incorporate numerically or
experimentally obtained spectral densities as well as an arbitrary range of
temperatures without any need of fitting procedures or numerically unfeasible
expansions. Secondly that it does so at no numerical penalty. This unique set of
properties of the C-HEOM approach renders it preferable to the original HEOM for
systems with low to intermediate coupling strengths. %

As the number of tiers needed to obtain converged exciton dynamics increases
with the system-bath coupling strength, higher coupling strengths can only be
dealt with using higher tiers of the hierarchy. It is possible, however, to
implement the C-HEOM on highly parallel processing units. This would allow the
C-HEOM to be used for higher coupling strengths and/or to obtain the dynamics
for longer simulation times. Complex spectral densities obtained from
experiments \cite{adol06a,kell13a} or from a combination of classical molecular
dynamics simulations and quantum chemical calculations  \cite{chan15a,jang18a}
 that show complex structures that
cannot be reproduced by sum of Lorentzian functions \cite{krei14a} may be
treated accurately using higher tiers of the C-HEOM approach. Work in this
direction is in progress.

 \section{Acknowledgments} We are grateful to Haobin Wang for providing the
ML-MCTDH data files. This work was supported by the Deutsche
Forschungsgemeinschaft through grants KL~1299/14-1 and  GRK~2247. 

\section{Appendix}

If the hierarchy is terminated at the level $n=0$ using the above approximation
on gets the usual second-order TL formalism\cite{klei04a,pope16a}. %
Setting $n$ to 2, which refers to second-tier truncation of the HEOM, and
using a two-site system ($j=2$) as an example, the zeroth tier reads
\begin{align}
\partial_t \bm{\rho}_{S}^{(0)} =& -i\mathcal{L}_{S}(t) \bm{\rho}_{S}^{(0)}\\\nonumber
& - \sum_{j_1}  K_{j_1}^{\times} \big( \sum_{k}^K (I_{j_1,k} \bm{\rho}_{j_{1},k}^{(1)}) - \sum_{k}^K (I_{j_1,k} \bm{\rho}_{j_{1},k}^{(1)} )^{\dagger}\big)\\\nonumber
\end{align}
while the first tier reads
\begin{align}
\partial_t \bm{\rho}_{j_{1},k}^{(1)} =& -i\mathcal{L}_{S}(t) \bm{\rho}_{j_{1},k}^{(1)}\\\nonumber
& - \sum_{j_2}  K_{j_2}^{\times} \big( \sum_{k}^K I_{j_2,k} \bm{\rho}_{j_{1}j_{2},k}^{(2)}  - \bm{\rho}_{j_{1},k}^{(1)} \Lambda_{j_{2}}^{\dagger}\big)\\\nonumber
& - i  \bar{\omega}\bm{\rho}_{j_{1},k}^{(1)} \\\nonumber
&+ \begin{cases}
 -\Omega \bm{\rho}_{j_{1},k}^{(1)} + K_{j_1} \bm{\rho}_{S}~~~~~~~~~~~~~~~~~~~\text{if}~ k = 0\\
  \frac{\Omega}{2} \big( \bm{\rho}_{j_{1},k-1}^{(1)} -  \bm{\rho}_{j_{1},k+1}^{(1)} \big)~~~~~~~~~~~~~~~\text{if}~0< k < K_{ch}\\
 \frac{\Omega}{2} \big( \bm{\rho}_{j_{1},k-1}^{(1)}\big)~~~~~~~~~~~~~~~~~~~~~~~~~~~~\text{if}~k = K_{ch}
\end{cases}~.
\end{align}\\
The TNL truncated second tier can be written as 
\begin{align}
\partial_t \bm{\rho}_{j_{1}j_{2},k}^{(2)} =& -i\mathcal{L}_{S}(t) \bm{\rho}_{j_{1}j_{2},k}^{(2)}\\\nonumber
& - \sum_2 i  \bar{\omega}\bm{\rho}_{j_{1}{j_2},k}^{(1)} \\\nonumber
&+ \begin{cases}
 -\sum_2 \Omega \bm{\rho}_{j_{1}j_{2},k}^{(1)}+ K_{j_1}
\bm{\rho}_{j_{2},k}^{(1)} + K_{j_2} \bm{\rho}_{j_{1},k}^{(1)}
~~~~~~~~~~~\text{if}~ k = 0\\ \sum_2 \frac{\Omega}{2} \big(
\bm{\rho}_{j_{1}j_{2},k-1}^{(1)} -  \bm{\rho}_{j_{1}j_{2},k+1}^{(1)}
\big)~~~~~~~~~~~~~~~~~~~~~~\text{if}~0< k < K_{ch}\\ \sum_2 \frac{\Omega}{2}
\big(
\bm{\rho}_{j_{1}j_{2},k-1}^{(1)}\big)~~~~~~~~~~~~~~~~~~~~~~~~~~~~~~~~~~~~~\text{if}~k = K_{ch}
\end{cases}~.
\end{align}

Note that one does not need $\bm{\rho}^{(n+)}$~ for the TL truncation and the full
fourth-order, TL4, results can be obtained with the
following equations
\begin{align}
\partial_t \bm{\rho}_{S}^{(0)} =& -i\mathcal{L}_{S}(t)
\bm{\rho}_{S}^{(0)}\\\nonumber & - \sum_{j_1}  K_{j_1}^{\times} \big( \sum_{k}^K
(I_{j_1,k} \bm{\rho}_{j_{1},k}^{(1)}) - \sum_{k}^K (I_{j_1,k}
\bm{\rho}_{j_{1},k}^{(1)})^{\dagger} \big)\\\nonumber
\end{align}
\begin{align}
\partial_t \bm{\rho}_{j_{1},k}^{(1)} =& -i\mathcal{L}_{S}(t) \bm{\rho}_{j_{1},k}^{(1)}\\\nonumber
& - \sum_{j_2}  K_{j_2}^{\times} \big( \Lambda_{j_2}\bm{\rho}_{j_{1},k}^{(1)} - \bm{\rho}_{j_{1},k}^{(1)} \Lambda_{j_{2}}^{\dagger}\big)\\\nonumber
& - i  \bar{\omega}\bm{\rho}_{j_{1},k}^{(1)} \\\nonumber
&+ \begin{cases}
 -\Omega \bm{\rho}_{j_{1},k}^{(1)} + K_{j_1} \bm{\rho}_{S}~~~~~~~~~~~~~~~~~~~\text{if}~ k = 0\\
  \frac{\Omega}{2} \big( \bm{\rho}_{j_{1},k-1}^{(1)} -  \bm{\rho}_{j_{1},k+1}^{(1)} \big)~~~~~~~~~~~~~~~\text{if}~0< k < K_{ch}\\
 \frac{\Omega}{2} \big( \bm{\rho}_{j_{1},k-1}^{(1)}\big)~~~~~~~~~~~~~~~~~~~~~~~~~~~\text{if}~k = K_{ch}
\end{cases}~.
\end{align}


\begin{thebibliography}{67}%
\makeatletter
\providecommand \@ifxundefined [1]{%
 \@ifx{#1\undefined}
}%
\providecommand \@ifnum [1]{%
 \ifnum #1\expandafter \@firstoftwo
 \else \expandafter \@secondoftwo
 \fi
}%
\providecommand \@ifx [1]{%
 \ifx #1\expandafter \@firstoftwo
 \else \expandafter \@secondoftwo
 \fi
}%
\providecommand \natexlab [1]{#1}%
\providecommand \enquote  [1]{``#1''}%
\providecommand \bibnamefont  [1]{#1}%
\providecommand \bibfnamefont [1]{#1}%
\providecommand \citenamefont [1]{#1}%
\providecommand \href@noop [0]{\@secondoftwo}%
\providecommand \href [0]{\begingroup \@sanitize@url \@href}%
\providecommand \@href[1]{\@@startlink{#1}\@@href}%
\providecommand \@@href[1]{\endgroup#1\@@endlink}%
\providecommand \@sanitize@url [0]{\catcode `\\12\catcode `\$12\catcode
  `\&12\catcode `\#12\catcode `\^12\catcode `\_12\catcode `\%12\relax}%
\providecommand \@@startlink[1]{}%
\providecommand \@@endlink[0]{}%
\providecommand \url  [0]{\begingroup\@sanitize@url \@url }%
\providecommand \@url [1]{\endgroup\@href {#1}{\urlprefix }}%
\providecommand \urlprefix  [0]{URL }%
\providecommand \Eprint [0]{\href }%
\providecommand \doibase [0]{http://dx.doi.org/}%
\providecommand \selectlanguage [0]{\@gobble}%
\providecommand \bibinfo  [0]{\@secondoftwo}%
\providecommand \bibfield  [0]{\@secondoftwo}%
\providecommand \translation [1]{[#1]}%
\providecommand \BibitemOpen [0]{}%
\providecommand \bibitemStop [0]{}%
\providecommand \bibitemNoStop [0]{.\EOS\space}%
\providecommand \EOS [0]{\spacefactor3000\relax}%
\providecommand \BibitemShut  [1]{\csname bibitem#1\endcsname}%
\let\auto@bib@innerbib\@empty
\bibitem [{\citenamefont {{W}eiss}(1999)}]{weis99}%
  \BibitemOpen
  \bibfield  {author} {\bibinfo {author} {\bibfnamefont {U.}~\bibnamefont
  {{W}eiss}},\ }\href@noop {} {\emph {\bibinfo {title} {{Q}uantum {D}issipative
  {S}ystems}}},\ \bibinfo {edition} {2nd}\ ed.\ (\bibinfo  {publisher} {World
  Scientific},\ \bibinfo {year} {1999})\BibitemShut {NoStop}%
\bibitem [{\citenamefont {{B}reuer}\ and\ \citenamefont
  {{P}etruccione}(2002)}]{breu02}%
  \BibitemOpen
  \bibfield  {author} {\bibinfo {author} {\bibfnamefont {H.~P.}\ \bibnamefont
  {{B}reuer}}\ and\ \bibinfo {author} {\bibfnamefont {F.}~\bibnamefont
  {{P}etruccione}},\ }\href@noop {} {\emph {\bibinfo {title} {{T}he {T}heory of
  {O}pen {Q}uantum {S}ystems}}}\ (\bibinfo  {publisher} {Oxford University
  Press},\ \bibinfo {year} {2002})\BibitemShut {NoStop}%
\bibitem [{\citenamefont {{M}ay}\ and\ \citenamefont {{K}\"uhn}(2011)}]{may11}%
  \BibitemOpen
  \bibfield  {author} {\bibinfo {author} {\bibfnamefont {V.}~\bibnamefont
  {{M}ay}}\ and\ \bibinfo {author} {\bibfnamefont {O.}~\bibnamefont
  {{K}\"uhn}},\ }\href@noop {} {\emph {\bibinfo {title} {{C}harge and {E}nergy
  {T}ransfer in {M}olecular {S}ystems}}},\ \bibinfo {edition} {3rd}\ ed.\
  (\bibinfo  {publisher} {Wiley--VCH},\ \bibinfo {year} {2011})\BibitemShut
  {NoStop}%
\bibitem [{\citenamefont {{T}animura}\ and\ \citenamefont
  {{K}ubo}(1989)}]{tani89}%
  \BibitemOpen
  \bibfield  {author} {\bibinfo {author} {\bibfnamefont {Y.}~\bibnamefont
  {{T}animura}}\ and\ \bibinfo {author} {\bibfnamefont {R.}~\bibnamefont
  {{K}ubo}},\ }\href {\doibase 10.1143/JPSJ.58.101} {\bibfield  {journal}
  {\bibinfo  {journal} {{J}. {P}hys. {S}oc. {J}pn.}\ }\textbf {\bibinfo
  {volume} {58}},\ \bibinfo {pages} {101} (\bibinfo {year} {1989})}\BibitemShut
  {NoStop}%
\bibitem [{\citenamefont {{I}shizaki}\ and\ \citenamefont
  {{F}leming}(2009{\natexlab{a}})}]{ishi09a}%
  \BibitemOpen
  \bibfield  {author} {\bibinfo {author} {\bibfnamefont {A.}~\bibnamefont
  {{I}shizaki}}\ and\ \bibinfo {author} {\bibfnamefont {G.~R.}\ \bibnamefont
  {{F}leming}},\ }\href {\doibase 10.1073/pnas.0908989106} {\bibfield
  {journal} {\bibinfo  {journal} {{P}roc. {N}atl. {A}cad. {S}ci. {USA}}\
  }\textbf {\bibinfo {volume} {106}},\ \bibinfo {pages} {17255} (\bibinfo
  {year} {2009}{\natexlab{a}})}\BibitemShut {NoStop}%
\bibitem [{\citenamefont {Prior}\ \emph {et~al.}(2010)\citenamefont {Prior},
  \citenamefont {Chin}, \citenamefont {Huelga},\ and\ \citenamefont
  {Plenio}}]{prio10a}%
  \BibitemOpen
  \bibfield  {author} {\bibinfo {author} {\bibfnamefont {J.}~\bibnamefont
  {Prior}}, \bibinfo {author} {\bibfnamefont {A.~W.}\ \bibnamefont {Chin}},
  \bibinfo {author} {\bibfnamefont {S.~F.}\ \bibnamefont {Huelga}}, \ and\
  \bibinfo {author} {\bibfnamefont {M.~B.}\ \bibnamefont {Plenio}},\ }\href
  {\doibase 10.1103/PhysRevLett.105.050404} {\bibfield  {journal} {\bibinfo
  {journal} {Phys. Rev. Lett.}\ }\textbf {\bibinfo {volume} {105}},\ \bibinfo
  {pages} {050404} (\bibinfo {year} {2010})}\BibitemShut {NoStop}%
\bibitem [{\citenamefont {{N}albach}, \citenamefont {{B}raun},\ and\
  \citenamefont {{T}horwart}(2011)}]{nalb11a}%
  \BibitemOpen
  \bibfield  {author} {\bibinfo {author} {\bibfnamefont {P.}~\bibnamefont
  {{N}albach}}, \bibinfo {author} {\bibfnamefont {D.}~\bibnamefont {{B}raun}},
  \ and\ \bibinfo {author} {\bibfnamefont {M.}~\bibnamefont {{T}horwart}},\
  }\href {\doibase 10.1103/physreve.84.041926} {\bibfield  {journal} {\bibinfo
  {journal} {{P}hys. {R}ev. {E}}\ }\textbf {\bibinfo {volume} {84}},\ \bibinfo
  {pages} {041926} (\bibinfo {year} {2011})}\BibitemShut {NoStop}%
\bibitem [{\citenamefont {{M}\"uhlbacher}\ and\ \citenamefont
  {{K}leinekath\"ofer}(2012)}]{muehl12a}%
  \BibitemOpen
  \bibfield  {author} {\bibinfo {author} {\bibfnamefont {L.}~\bibnamefont
  {{M}\"uhlbacher}}\ and\ \bibinfo {author} {\bibfnamefont {U.}~\bibnamefont
  {{K}leinekath\"ofer}},\ }\href {\doibase 10.1021/jp301444q} {\bibfield
  {journal} {\bibinfo  {journal} {{J}. {P}hys. {C}hem. {B}}\ }\textbf {\bibinfo
  {volume} {116}},\ \bibinfo {pages} {3900} (\bibinfo {year}
  {2012})}\BibitemShut {NoStop}%
\bibitem [{\citenamefont {Schr{\"{o}}ter}\ \emph {et~al.}(2015)\citenamefont
  {Schr{\"{o}}ter}, \citenamefont {Ivanov}, \citenamefont {Schulze},
  \citenamefont {Polyutov}, \citenamefont {Yan}, \citenamefont {Pullerits},\
  and\ \citenamefont {K{\"{u}}hn}}]{schr15c}%
  \BibitemOpen
  \bibfield  {author} {\bibinfo {author} {\bibfnamefont {M.}~\bibnamefont
  {Schr{\"{o}}ter}}, \bibinfo {author} {\bibfnamefont {S.~D.}\ \bibnamefont
  {Ivanov}}, \bibinfo {author} {\bibfnamefont {J.}~\bibnamefont {Schulze}},
  \bibinfo {author} {\bibfnamefont {S.~P.}\ \bibnamefont {Polyutov}}, \bibinfo
  {author} {\bibfnamefont {Y.}~\bibnamefont {Yan}}, \bibinfo {author}
  {\bibfnamefont {T.}~\bibnamefont {Pullerits}}, \ and\ \bibinfo {author}
  {\bibfnamefont {O.}~\bibnamefont {K{\"{u}}hn}},\ }\href {\doibase
  10.1016/j.physrep.2014.12.001} {\bibfield  {journal} {\bibinfo  {journal}
  {Phys. Rep.}\ }\textbf {\bibinfo {volume} {567}},\ \bibinfo {pages} {1}
  (\bibinfo {year} {2015})}\BibitemShut {NoStop}%
\bibitem [{\citenamefont {{X}u}\ and\ \citenamefont {{Y}an}(2007)}]{xu07a}%
  \BibitemOpen
  \bibfield  {author} {\bibinfo {author} {\bibfnamefont {R.-X.}\ \bibnamefont
  {{X}u}}\ and\ \bibinfo {author} {\bibfnamefont {Y.}~\bibnamefont {{Y}an}},\
  }\href@noop {} {\bibfield  {journal} {\bibinfo  {journal} {{P}hys. {R}ev.
  {E}}\ }\textbf {\bibinfo {volume} {75}},\ \bibinfo {pages} {031107} (\bibinfo
  {year} {2007})}\BibitemShut {NoStop}%
\bibitem [{\citenamefont {{X}u}\ \emph {et~al.}(2004)\citenamefont {{X}u},
  \citenamefont {{C}ui}, \citenamefont {{L}i}, \citenamefont {{M}o},\ and\
  \citenamefont {{Y}an}}]{xu04b}%
  \BibitemOpen
  \bibfield  {author} {\bibinfo {author} {\bibfnamefont {R.-X.}\ \bibnamefont
  {{X}u}}, \bibinfo {author} {\bibfnamefont {P.}~\bibnamefont {{C}ui}},
  \bibinfo {author} {\bibfnamefont {X.-Q.}\ \bibnamefont {{L}i}}, \bibinfo
  {author} {\bibfnamefont {Y.}~\bibnamefont {{M}o}}, \ and\ \bibinfo {author}
  {\bibfnamefont {Y.~J.}\ \bibnamefont {{Y}an}},\ }\href@noop {} {\bibfield
  {journal} {\bibinfo  {journal} {{J}. {C}hem. {P}hys.}\ }\textbf {\bibinfo
  {volume} {122}},\ \bibinfo {pages} {041103} (\bibinfo {year}
  {2004})}\BibitemShut {NoStop}%
\bibitem [{\citenamefont {{T}animura}(2006)}]{tani06}%
  \BibitemOpen
  \bibfield  {author} {\bibinfo {author} {\bibfnamefont {Y.}~\bibnamefont
  {{T}animura}},\ }\href {\doibase 10.1143/JPSJ.75.082001} {\bibfield
  {journal} {\bibinfo  {journal} {{J}. {P}hys. {S}oc. {J}pn.}\ }\textbf
  {\bibinfo {volume} {75}},\ \bibinfo {pages} {082001} (\bibinfo {year}
  {2006})}\BibitemShut {NoStop}%
\bibitem [{\citenamefont {{Y}an}, \citenamefont {{Y}ang},\ and\ \citenamefont
  {{S}hao}(2004)}]{yan04}%
  \BibitemOpen
  \bibfield  {author} {\bibinfo {author} {\bibfnamefont {Y.~J.}\ \bibnamefont
  {{Y}an}}, \bibinfo {author} {\bibfnamefont {F.}~\bibnamefont {{Y}ang}}, \
  and\ \bibinfo {author} {\bibfnamefont {J.}~\bibnamefont {{S}hao}},\
  }\href@noop {} {\bibfield  {journal} {\bibinfo  {journal} {{C}hem. {P}hys.
  {L}ett.}\ }\textbf {\bibinfo {volume} {395}},\ \bibinfo {pages} {216}
  (\bibinfo {year} {2004})}\BibitemShut {NoStop}%
\bibitem [{\citenamefont {Shao}(2004)}]{shao04a}%
  \BibitemOpen
  \bibfield  {author} {\bibinfo {author} {\bibfnamefont {J.}~\bibnamefont
  {Shao}},\ }\href {\doibase 10.1063/1.1647528} {\bibfield  {journal} {\bibinfo
   {journal} {J. Chem. Phys.}\ }\textbf {\bibinfo {volume} {120}},\ \bibinfo
  {pages} {5053} (\bibinfo {year} {2004})}\BibitemShut {NoStop}%
\bibitem [{\citenamefont {{I}shizaki}\ and\ \citenamefont
  {{T}animura}(2005)}]{ishi05a}%
  \BibitemOpen
  \bibfield  {author} {\bibinfo {author} {\bibfnamefont {A.}~\bibnamefont
  {{I}shizaki}}\ and\ \bibinfo {author} {\bibfnamefont {Y.}~\bibnamefont
  {{T}animura}},\ }\href {\doibase 10.1143/JPSJ.74.3131} {\bibfield  {journal}
  {\bibinfo  {journal} {{J}. {P}hys. {S}oc. {J}pn.}\ }\textbf {\bibinfo
  {volume} {74}},\ \bibinfo {pages} {3131} (\bibinfo {year}
  {2005})}\BibitemShut {NoStop}%
\bibitem [{\citenamefont {{A}min}\ \emph {et~al.}(2009)\citenamefont {{A}min},
  \citenamefont {{L}i}, \citenamefont {{P}hillips},\ and\ \citenamefont
  {{K}leinekath\"ofer}}]{amin09a}%
  \BibitemOpen
  \bibfield  {author} {\bibinfo {author} {\bibfnamefont {A.~F.}\ \bibnamefont
  {{A}min}}, \bibinfo {author} {\bibfnamefont {G.-Q.}\ \bibnamefont {{L}i}},
  \bibinfo {author} {\bibfnamefont {A.~H.}\ \bibnamefont {{P}hillips}}, \ and\
  \bibinfo {author} {\bibfnamefont {U.}~\bibnamefont {{K}leinekath\"ofer}},\
  }\href {\doibase 10.1140/epjb/e2009-00075-9} {\bibfield  {journal} {\bibinfo
  {journal} {{E}ur. {P}hys. {J}. {B}}\ }\textbf {\bibinfo {volume} {68}},\
  \bibinfo {pages} {103} (\bibinfo {year} {2009})}\BibitemShut {NoStop}%
\bibitem [{\citenamefont {{I}shizaki}\ and\ \citenamefont
  {{F}leming}(2009{\natexlab{b}})}]{ishi09b}%
  \BibitemOpen
  \bibfield  {author} {\bibinfo {author} {\bibfnamefont {A.}~\bibnamefont
  {{I}shizaki}}\ and\ \bibinfo {author} {\bibfnamefont {G.~R.}\ \bibnamefont
  {{F}leming}},\ }\href {\doibase 10.1063/1.3155372} {\bibfield  {journal}
  {\bibinfo  {journal} {{J}. {C}hem. {P}hys.}\ }\textbf {\bibinfo {volume}
  {130}},\ \bibinfo {eid} {234111} (\bibinfo {year}
  {2009}{\natexlab{b}})}\BibitemShut {NoStop}%
\bibitem [{\citenamefont {{K}reisbeck}\ \emph {et~al.}(2011)\citenamefont
  {{K}reisbeck}, \citenamefont {{K}ramer}, \citenamefont {{R}odriguez},\ and\
  \citenamefont {{H}ein}}]{krei11a}%
  \BibitemOpen
  \bibfield  {author} {\bibinfo {author} {\bibfnamefont {C.}~\bibnamefont
  {{K}reisbeck}}, \bibinfo {author} {\bibfnamefont {T.}~\bibnamefont
  {{K}ramer}}, \bibinfo {author} {\bibfnamefont {M.}~\bibnamefont
  {{R}odriguez}}, \ and\ \bibinfo {author} {\bibfnamefont {B.}~\bibnamefont
  {{H}ein}},\ }\href {\doibase 10.1021/ct200126d} {\bibfield  {journal}
  {\bibinfo  {journal} {{J}. {C}hem. {T}heory {C}omput.}\ }\textbf {\bibinfo
  {volume} {7}},\ \bibinfo {pages} {2166} (\bibinfo {year} {2011})}\BibitemShut
  {NoStop}%
\bibitem [{\citenamefont {{S}tr\"umpfer}\ and\ \citenamefont
  {{S}chulten}(2012{\natexlab{a}})}]{stru12b}%
  \BibitemOpen
  \bibfield  {author} {\bibinfo {author} {\bibfnamefont {J.}~\bibnamefont
  {{S}tr\"umpfer}}\ and\ \bibinfo {author} {\bibfnamefont {K.}~\bibnamefont
  {{S}chulten}},\ }\href {\doibase 10.1021/ct3003833} {\bibfield  {journal}
  {\bibinfo  {journal} {{J}. {C}hem. {T}heory {C}omput.}\ }\textbf {\bibinfo
  {volume} {8}},\ \bibinfo {pages} {2808} (\bibinfo {year}
  {2012}{\natexlab{a}})}\BibitemShut {NoStop}%
\bibitem [{\citenamefont {Kramer}\ \emph {et~al.}(2018)\citenamefont {Kramer},
  \citenamefont {Noack}, \citenamefont {Reinefeld}, \citenamefont
  {Rodr{\'{i}}guez},\ and\ \citenamefont {Zelinskyy}}]{kram18a}%
  \BibitemOpen
  \bibfield  {author} {\bibinfo {author} {\bibfnamefont {T.}~\bibnamefont
  {Kramer}}, \bibinfo {author} {\bibfnamefont {M.}~\bibnamefont {Noack}},
  \bibinfo {author} {\bibfnamefont {A.}~\bibnamefont {Reinefeld}}, \bibinfo
  {author} {\bibfnamefont {M.}~\bibnamefont {Rodr{\'{i}}guez}}, \ and\ \bibinfo
  {author} {\bibfnamefont {Y.}~\bibnamefont {Zelinskyy}},\ }\href {\doibase
  10.1002/jcc.25354} {\bibfield  {journal} {\bibinfo  {journal} {J. Comput.
  Chem.}\ }\textbf {\bibinfo {volume} {39}},\ \bibinfo {pages} {1779} (\bibinfo
  {year} {2018})}\BibitemShut {NoStop}%
\bibitem [{\citenamefont {{W}ilkins}\ and\ \citenamefont
  {{D}attani}(2015)}]{wilk15a}%
  \BibitemOpen
  \bibfield  {author} {\bibinfo {author} {\bibfnamefont {D.~M.}\ \bibnamefont
  {{W}ilkins}}\ and\ \bibinfo {author} {\bibfnamefont {N.~S.}\ \bibnamefont
  {{D}attani}},\ }\href {\doibase 10.1021/ct501066k} {\bibfield  {journal}
  {\bibinfo  {journal} {{J}. {C}hem. {T}heory {C}omput.}\ }\textbf {\bibinfo
  {volume} {11}},\ \bibinfo {pages} {3411} (\bibinfo {year}
  {2015})}\BibitemShut {NoStop}%
\bibitem [{\citenamefont {{S}tr\"umpfer}\ and\ \citenamefont
  {{S}chulten}(2012{\natexlab{b}})}]{stru12c}%
  \BibitemOpen
  \bibfield  {author} {\bibinfo {author} {\bibfnamefont {J.}~\bibnamefont
  {{S}tr\"umpfer}}\ and\ \bibinfo {author} {\bibfnamefont {K.}~\bibnamefont
  {{S}chulten}},\ }\href {\doibase 10.1063/1.4738953} {\bibfield  {journal}
  {\bibinfo  {journal} {{J}. {C}hem. {P}hys.}\ }\textbf {\bibinfo {volume}
  {137}},\ \bibinfo {pages} {065101} (\bibinfo {year}
  {2012}{\natexlab{b}})}\BibitemShut {NoStop}%
\bibitem [{\citenamefont {{L}iu}\ \emph {et~al.}(2014)\citenamefont {{L}iu},
  \citenamefont {{Z}hu}, \citenamefont {{B}ai},\ and\ \citenamefont
  {{S}hi}}]{liu14a}%
  \BibitemOpen
  \bibfield  {author} {\bibinfo {author} {\bibfnamefont {H.}~\bibnamefont
  {{L}iu}}, \bibinfo {author} {\bibfnamefont {L.}~\bibnamefont {{Z}hu}},
  \bibinfo {author} {\bibfnamefont {S.}~\bibnamefont {{B}ai}}, \ and\ \bibinfo
  {author} {\bibfnamefont {Q.}~\bibnamefont {{S}hi}},\ }\href {\doibase
  10.1063/1.4870035} {\bibfield  {journal} {\bibinfo  {journal} {{J}. {C}hem.
  {P}hys.}\ }\textbf {\bibinfo {volume} {140}},\ \bibinfo {eid} {134106}
  (\bibinfo {year} {2014})}\BibitemShut {NoStop}%
\bibitem [{\citenamefont {{M}eier}\ and\ \citenamefont
  {{T}annor}(1999)}]{meie99}%
  \BibitemOpen
  \bibfield  {author} {\bibinfo {author} {\bibfnamefont {C.}~\bibnamefont
  {{M}eier}}\ and\ \bibinfo {author} {\bibfnamefont {D.~J.}\ \bibnamefont
  {{T}annor}},\ }\href {\doibase 10.1063/1.479669} {\bibfield  {journal}
  {\bibinfo  {journal} {{J}. {C}hem. {P}hys.}\ }\textbf {\bibinfo {volume}
  {111}},\ \bibinfo {pages} {3365} (\bibinfo {year} {1999})}\BibitemShut
  {NoStop}%
\bibitem [{\citenamefont {{R}itschel}\ and\ \citenamefont
  {{E}isfeld}(2014)}]{rits14a}%
  \BibitemOpen
  \bibfield  {author} {\bibinfo {author} {\bibfnamefont {G.}~\bibnamefont
  {{R}itschel}}\ and\ \bibinfo {author} {\bibfnamefont {A.}~\bibnamefont
  {{E}isfeld}},\ }\href {\doibase 10.1063/1.4893931} {\bibfield  {journal}
  {\bibinfo  {journal} {{J}. {C}hem. {P}hys.}\ }\textbf {\bibinfo {volume}
  {141}},\ \bibinfo {pages} {094101} (\bibinfo {year} {2014})}\BibitemShut
  {NoStop}%
\bibitem [{\citenamefont {{C}handrasekaran}\ \emph {et~al.}(2015)\citenamefont
  {{C}handrasekaran}, \citenamefont {{A}ghtar}, \citenamefont {{V}alleau},
  \citenamefont {Aspuru-Guzik},\ and\ \citenamefont
  {{K}leinekath\"ofer}}]{chan15a}%
  \BibitemOpen
  \bibfield  {author} {\bibinfo {author} {\bibfnamefont {S.}~\bibnamefont
  {{C}handrasekaran}}, \bibinfo {author} {\bibfnamefont {M.}~\bibnamefont
  {{A}ghtar}}, \bibinfo {author} {\bibfnamefont {S.}~\bibnamefont {{V}alleau}},
  \bibinfo {author} {\bibfnamefont {A.}~\bibnamefont {Aspuru-Guzik}}, \ and\
  \bibinfo {author} {\bibfnamefont {U.}~\bibnamefont {{K}leinekath\"ofer}},\
  }\href {\doibase 10.1021/acs.jpcb.5b03654} {\bibfield  {journal} {\bibinfo
  {journal} {{J}. {P}hys. {C}hem. {B}}\ }\textbf {\bibinfo {volume} {119}},\
  \bibinfo {pages} {9995} (\bibinfo {year} {2015})}\BibitemShut {NoStop}%
\bibitem [{\citenamefont {Jang}\ and\ \citenamefont
  {Mennucci}(2018)}]{jang18a}%
  \BibitemOpen
  \bibfield  {author} {\bibinfo {author} {\bibfnamefont {S.~J.}\ \bibnamefont
  {Jang}}\ and\ \bibinfo {author} {\bibfnamefont {B.}~\bibnamefont
  {Mennucci}},\ }\href {\doibase 10.1103/RevModPhys.90.035003} {\bibfield
  {journal} {\bibinfo  {journal} {Rev. Mod. Phys.}\ }\textbf {\bibinfo {volume}
  {90}},\ \bibinfo {pages} {035003} (\bibinfo {year} {2018})}\BibitemShut
  {NoStop}%
\bibitem [{\citenamefont {Dattani}, \citenamefont {Pollock},\ and\
  \citenamefont {Wilkins}(2012)}]{datt12a}%
  \BibitemOpen
  \bibfield  {author} {\bibinfo {author} {\bibfnamefont {N.~S.}\ \bibnamefont
  {Dattani}}, \bibinfo {author} {\bibfnamefont {F.}~\bibnamefont {Pollock}}, \
  and\ \bibinfo {author} {\bibfnamefont {D.~M.}\ \bibnamefont {Wilkins}},\
  }\href {http://www.naturalspublishing.com/files/published/464k51t1luip94.pdf}
  {\bibfield  {journal} {\bibinfo  {journal} {Quant. Phys. Lett.}\ }\textbf
  {\bibinfo {volume} {1}},\ \bibinfo {pages} {35} (\bibinfo {year}
  {2012})}\BibitemShut {NoStop}%
\bibitem [{\citenamefont {{T}ang}\ \emph {et~al.}(2015)\citenamefont {{T}ang},
  \citenamefont {{O}uyang}, \citenamefont {{G}ong}, \citenamefont {{W}ang},\
  and\ \citenamefont {{W}u}}]{tang15a}%
  \BibitemOpen
  \bibfield  {author} {\bibinfo {author} {\bibfnamefont {Z.}~\bibnamefont
  {{T}ang}}, \bibinfo {author} {\bibfnamefont {X.}~\bibnamefont {{O}uyang}},
  \bibinfo {author} {\bibfnamefont {Z.}~\bibnamefont {{G}ong}}, \bibinfo
  {author} {\bibfnamefont {H.}~\bibnamefont {{W}ang}}, \ and\ \bibinfo {author}
  {\bibfnamefont {J.}~\bibnamefont {{W}u}},\ }\href {\doibase
  10.1063/1.4936924} {\bibfield  {journal} {\bibinfo  {journal} {{J}. {C}hem.
  {P}hys.}\ }\textbf {\bibinfo {volume} {143}},\ \bibinfo {eid} {224112}
  (\bibinfo {year} {2015}),\ 10.1063/1.4936924}\BibitemShut {NoStop}%
\bibitem [{\citenamefont {{P}opescu}, \citenamefont {{R}ahman},\ and\
  \citenamefont {{K}leinekath\"ofer}(2015)}]{pope15a}%
  \BibitemOpen
  \bibfield  {author} {\bibinfo {author} {\bibfnamefont {B.}~\bibnamefont
  {{P}opescu}}, \bibinfo {author} {\bibfnamefont {H.}~\bibnamefont {{R}ahman}},
  \ and\ \bibinfo {author} {\bibfnamefont {U.}~\bibnamefont
  {{K}leinekath\"ofer}},\ }\href {\doibase 10.1063/1.4917198} {\bibfield
  {journal} {\bibinfo  {journal} {{J}. {C}hem. {P}hys.}\ }\textbf {\bibinfo
  {volume} {142}},\ \bibinfo {eid} {154103} (\bibinfo {year}
  {2015})}\BibitemShut {NoStop}%
\bibitem [{\citenamefont {{P}opescu}, \citenamefont {{R}ahman},\ and\
  \citenamefont {{K}leinekath\"ofer}(2016)}]{pope16a}%
  \BibitemOpen
  \bibfield  {author} {\bibinfo {author} {\bibfnamefont {B.}~\bibnamefont
  {{P}opescu}}, \bibinfo {author} {\bibfnamefont {H.}~\bibnamefont {{R}ahman}},
  \ and\ \bibinfo {author} {\bibfnamefont {U.}~\bibnamefont
  {{K}leinekath\"ofer}},\ }\href {\doibase 10.1021/acs.jpca.5b12237} {\bibfield
   {journal} {\bibinfo  {journal} {J. Phys. Chem. A}\ }\textbf {\bibinfo
  {volume} {120}},\ \bibinfo {pages} {3270} (\bibinfo {year}
  {2016})}\BibitemShut {NoStop}%
\bibitem [{\citenamefont {Rahman}\ and\ \citenamefont
  {Kleinekath{\"{o}}fer}(2018)}]{rahm18a}%
  \BibitemOpen
  \bibfield  {author} {\bibinfo {author} {\bibfnamefont {H.}~\bibnamefont
  {Rahman}}\ and\ \bibinfo {author} {\bibfnamefont {U.}~\bibnamefont
  {Kleinekath{\"{o}}fer}},\ }\href {\doibase 10.1063/1.5054312} {\bibfield
  {journal} {\bibinfo  {journal} {J. Chem. Phys.}\ }\textbf {\bibinfo {volume}
  {149}},\ \bibinfo {pages} {234108} (\bibinfo {year} {2018})}\BibitemShut
  {NoStop}%
\bibitem [{\citenamefont {{T}ian}\ and\ \citenamefont
  {{C}hen}(2013)}]{tian13a}%
  \BibitemOpen
  \bibfield  {author} {\bibinfo {author} {\bibfnamefont {H.}~\bibnamefont
  {{T}ian}}\ and\ \bibinfo {author} {\bibfnamefont {G.}~\bibnamefont
  {{C}hen}},\ }\href {\doibase 10.1140/epjb/e2013-40333-7} {\bibfield
  {journal} {\bibinfo  {journal} {{E}ur. {P}hys. {J}. {B}}\ }\textbf {\bibinfo
  {volume} {86}},\ \bibinfo {eid} {411} (\bibinfo {year} {2013})}\BibitemShut
  {NoStop}%
\bibitem [{\citenamefont {{C}roy}\ and\ \citenamefont
  {{S}aalmann}(2009)}]{croy09a}%
  \BibitemOpen
  \bibfield  {author} {\bibinfo {author} {\bibfnamefont {A.}~\bibnamefont
  {{C}roy}}\ and\ \bibinfo {author} {\bibfnamefont {U.}~\bibnamefont
  {{S}aalmann}},\ }\href {\doibase 10.1103/PhysRevB.80.245311} {\bibfield
  {journal} {\bibinfo  {journal} {{P}hys. {R}ev. {B}}\ }\textbf {\bibinfo
  {volume} {80}},\ \bibinfo {pages} {245311} (\bibinfo {year}
  {2009})}\BibitemShut {NoStop}%
\bibitem [{\citenamefont {Erpenbeck}\ \emph {et~al.}(2018)\citenamefont
  {Erpenbeck}, \citenamefont {Hertlein}, \citenamefont {Schinabeck},\ and\
  \citenamefont {Thoss}}]{erpe18a}%
  \BibitemOpen
  \bibfield  {author} {\bibinfo {author} {\bibfnamefont {A.}~\bibnamefont
  {Erpenbeck}}, \bibinfo {author} {\bibfnamefont {C.}~\bibnamefont {Hertlein}},
  \bibinfo {author} {\bibfnamefont {C.}~\bibnamefont {Schinabeck}}, \ and\
  \bibinfo {author} {\bibfnamefont {M.}~\bibnamefont {Thoss}},\ }\href
  {\doibase 10.1063/1.5041716} {\bibfield  {journal} {\bibinfo  {journal} {J.
  Chem.Phys.}\ }\textbf {\bibinfo {volume} {149}},\ \bibinfo {pages} {064106}
  (\bibinfo {year} {2018})}\BibitemShut {NoStop}%
\bibitem [{\citenamefont {{K}osloff}(1994)}]{kosl94}%
  \BibitemOpen
  \bibfield  {author} {\bibinfo {author} {\bibfnamefont {R.}~\bibnamefont
  {{K}osloff}},\ }\href {\doibase 10.1146/annurev.pc.45.100194.001045}
  {\bibfield  {journal} {\bibinfo  {journal} {{A}nnu. {R}ev. {P}hys. {C}hem.}\
  }\textbf {\bibinfo {volume} {45}},\ \bibinfo {pages} {145} (\bibinfo {year}
  {1994})}\BibitemShut {NoStop}%
\bibitem [{\citenamefont {{N}dong}\ \emph {et~al.}(2009)\citenamefont
  {{N}dong}, \citenamefont {{T}al {E}zer}, \citenamefont {{K}osloff},\ and\
  \citenamefont {{K}och}}]{ndon09a}%
  \BibitemOpen
  \bibfield  {author} {\bibinfo {author} {\bibfnamefont {M.}~\bibnamefont
  {{N}dong}}, \bibinfo {author} {\bibfnamefont {H.}~\bibnamefont {{T}al
  {E}zer}}, \bibinfo {author} {\bibfnamefont {R.}~\bibnamefont {{K}osloff}}, \
  and\ \bibinfo {author} {\bibfnamefont {C.~P.}\ \bibnamefont {{K}och}},\
  }\href {\doibase 10.1063/1.3098940} {\bibfield  {journal} {\bibinfo
  {journal} {{J}. {C}hem. {P}hys.}\ }\textbf {\bibinfo {volume} {130}},\
  \bibinfo {eid} {124108} (\bibinfo {year} {2009})}\BibitemShut {NoStop}%
\bibitem [{\citenamefont {{N}dong}\ \emph {et~al.}(2010)\citenamefont
  {{N}dong}, \citenamefont {{T}al {E}zer}, \citenamefont {{K}osloff},\ and\
  \citenamefont {{K}och}}]{ndon10a}%
  \BibitemOpen
  \bibfield  {author} {\bibinfo {author} {\bibfnamefont {M.}~\bibnamefont
  {{N}dong}}, \bibinfo {author} {\bibfnamefont {H.}~\bibnamefont {{T}al
  {E}zer}}, \bibinfo {author} {\bibfnamefont {R.}~\bibnamefont {{K}osloff}}, \
  and\ \bibinfo {author} {\bibfnamefont {C.~P.}\ \bibnamefont {{K}och}},\
  }\href {\doibase 10.1063/1.3312531} {\bibfield  {journal} {\bibinfo
  {journal} {{J}. {C}hem. {P}hys.}\ }\textbf {\bibinfo {volume} {132}},\
  \bibinfo {eid} {064105} (\bibinfo {year} {2010})}\BibitemShut {NoStop}%
\bibitem [{\citenamefont {{G}uo}\ and\ \citenamefont {{C}hen}(1999)}]{guo99}%
  \BibitemOpen
  \bibfield  {author} {\bibinfo {author} {\bibfnamefont {H.}~\bibnamefont
  {{G}uo}}\ and\ \bibinfo {author} {\bibfnamefont {R.}~\bibnamefont {{C}hen}},\
  }\href {\doibase 10.1063/1.478570} {\bibfield  {journal} {\bibinfo  {journal}
  {{J}. {C}hem. {P}hys.}\ }\textbf {\bibinfo {volume} {110}},\ \bibinfo {pages}
  {6626} (\bibinfo {year} {1999})}\BibitemShut {NoStop}%
\bibitem [{\citenamefont {{H}uisinga}\ \emph {et~al.}(1999)\citenamefont
  {{H}uisinga}, \citenamefont {{P}esce}, \citenamefont {{K}osloff},\ and\
  \citenamefont {{S}aalfrank}}]{huis99}%
  \BibitemOpen
  \bibfield  {author} {\bibinfo {author} {\bibfnamefont {W.}~\bibnamefont
  {{H}uisinga}}, \bibinfo {author} {\bibfnamefont {L.}~\bibnamefont {{P}esce}},
  \bibinfo {author} {\bibfnamefont {R.}~\bibnamefont {{K}osloff}}, \ and\
  \bibinfo {author} {\bibfnamefont {P.}~\bibnamefont {{S}aalfrank}},\ }\href
  {\doibase 10.1063/1.478451} {\bibfield  {journal} {\bibinfo  {journal} {{J}.
  {C}hem. {P}hys.}\ }\textbf {\bibinfo {volume} {110}},\ \bibinfo {pages}
  {5538} (\bibinfo {year} {1999})}\BibitemShut {NoStop}%
\bibitem [{\citenamefont {{W}oods}\ \emph {et~al.}(2014)\citenamefont
  {{W}oods}, \citenamefont {{G}roux}, \citenamefont {{C}hin}, \citenamefont
  {{H}uelga},\ and\ \citenamefont {{P}lenio}}]{wood14a}%
  \BibitemOpen
  \bibfield  {author} {\bibinfo {author} {\bibfnamefont {M.~P.}\ \bibnamefont
  {{W}oods}}, \bibinfo {author} {\bibfnamefont {R.}~\bibnamefont {{G}roux}},
  \bibinfo {author} {\bibfnamefont {A.~W.}\ \bibnamefont {{C}hin}}, \bibinfo
  {author} {\bibfnamefont {S.~F.}\ \bibnamefont {{H}uelga}}, \ and\ \bibinfo
  {author} {\bibfnamefont {M.~B.}\ \bibnamefont {{P}lenio}},\ }\href {\doibase
  10.1063/1.4866769} {\bibfield  {journal} {\bibinfo  {journal} {{J}. {M}ath.
  {P}hys.}\ }\textbf {\bibinfo {volume} {55}},\ \bibinfo {eid} {032101}
  (\bibinfo {year} {2014})}\BibitemShut {NoStop}%
\bibitem [{\citenamefont {Suess}, \citenamefont {Eisfeld},\ and\ \citenamefont
  {Strunz}(2014)}]{sues14a}%
  \BibitemOpen
  \bibfield  {author} {\bibinfo {author} {\bibfnamefont {D.}~\bibnamefont
  {Suess}}, \bibinfo {author} {\bibfnamefont {A.}~\bibnamefont {Eisfeld}}, \
  and\ \bibinfo {author} {\bibfnamefont {W.~T.}\ \bibnamefont {Strunz}},\
  }\href {\doibase 10.1103/physrevlett.113.150403} {\bibfield  {journal}
  {\bibinfo  {journal} {Phys. Rev. Lett.}\ }\textbf {\bibinfo {volume} {113}}
  (\bibinfo {year} {2014}),\ 10.1103/physrevlett.113.150403}\BibitemShut
  {NoStop}%
\bibitem [{\citenamefont {{W}ang}\ and\ \citenamefont
  {{T}hoss}(2003)}]{wang03}%
  \BibitemOpen
  \bibfield  {author} {\bibinfo {author} {\bibfnamefont {H.}~\bibnamefont
  {{W}ang}}\ and\ \bibinfo {author} {\bibfnamefont {M.}~\bibnamefont
  {{T}hoss}},\ }\href@noop {} {\bibfield  {journal} {\bibinfo  {journal} {{J}.
  {C}hem. {P}hys.}\ }\textbf {\bibinfo {volume} {119}},\ \bibinfo {pages}
  {1289} (\bibinfo {year} {2003})}\BibitemShut {NoStop}%
\bibitem [{\citenamefont {Wang}\ and\ \citenamefont {Thoss}(2008)}]{wang08b}%
  \BibitemOpen
  \bibfield  {author} {\bibinfo {author} {\bibfnamefont {H.}~\bibnamefont
  {Wang}}\ and\ \bibinfo {author} {\bibfnamefont {M.}~\bibnamefont {Thoss}},\
  }\href {\doibase 10.1088/1367-2630/10/11/115005} {\bibfield  {journal}
  {\bibinfo  {journal} {New J. Phys.}\ }\textbf {\bibinfo {volume} {10}},\
  \bibinfo {pages} {115005} (\bibinfo {year} {2008})}\BibitemShut {NoStop}%
\bibitem [{\citenamefont {{F}eynman}\ and\ \citenamefont
  {{H}ibbs}(1965)}]{feyn65}%
  \BibitemOpen
  \bibfield  {author} {\bibinfo {author} {\bibfnamefont {R.~P.}\ \bibnamefont
  {{F}eynman}}\ and\ \bibinfo {author} {\bibfnamefont {A.~R.}\ \bibnamefont
  {{H}ibbs}},\ }\href@noop {} {\emph {\bibinfo {title} {{Q}uantum {M}echanics
  and {P}ath {I}ntegrals}}}\ (\bibinfo  {publisher} {McGraw-Hill},\ \bibinfo
  {year} {1965})\BibitemShut {NoStop}%
\bibitem [{\citenamefont {{N}albach}, \citenamefont {{E}ckel},\ and\
  \citenamefont {{T}horwart}(2010)}]{nalb10a}%
  \BibitemOpen
  \bibfield  {author} {\bibinfo {author} {\bibfnamefont {P.}~\bibnamefont
  {{N}albach}}, \bibinfo {author} {\bibfnamefont {J.}~\bibnamefont {{E}ckel}},
  \ and\ \bibinfo {author} {\bibfnamefont {M.}~\bibnamefont {{T}horwart}},\
  }\href {\doibase 10.1088/1367-2630/12/6/065043} {\bibfield  {journal}
  {\bibinfo  {journal} {{N}ew {J}. {P}hys.}\ }\textbf {\bibinfo {volume}
  {12}},\ \bibinfo {pages} {065043} (\bibinfo {year} {2010})}\BibitemShut
  {NoStop}%
\bibitem [{\citenamefont {Moix}, \citenamefont {Zhao},\ and\ \citenamefont
  {Cao}(2012)}]{moix12a}%
  \BibitemOpen
  \bibfield  {author} {\bibinfo {author} {\bibfnamefont {J.~M.}\ \bibnamefont
  {Moix}}, \bibinfo {author} {\bibfnamefont {Y.}~\bibnamefont {Zhao}}, \ and\
  \bibinfo {author} {\bibfnamefont {J.}~\bibnamefont {Cao}},\ }\href {\doibase
  10.1103/PhysRevB.85.115412} {\bibfield  {journal} {\bibinfo  {journal} {Phys.
  Rev. B}\ }\textbf {\bibinfo {volume} {85}},\ \bibinfo {pages} {115412}
  (\bibinfo {year} {2012})}\BibitemShut {NoStop}%
\bibitem [{\citenamefont {{K}leinekath\"ofer}(2004)}]{klei04a}%
  \BibitemOpen
  \bibfield  {author} {\bibinfo {author} {\bibfnamefont {U.}~\bibnamefont
  {{K}leinekath\"ofer}},\ }\href {\doibase 10.1063/1.1770619} {\bibfield
  {journal} {\bibinfo  {journal} {{J}. {C}hem. {P}hys.}\ }\textbf {\bibinfo
  {volume} {121}},\ \bibinfo {pages} {2505} (\bibinfo {year}
  {2004})}\BibitemShut {NoStop}%
\bibitem [{\citenamefont {{A}rfken}(1985)}]{arfk85}%
  \BibitemOpen
  \bibfield  {author} {\bibinfo {author} {\bibfnamefont {G.}~\bibnamefont
  {{A}rfken}},\ }\href@noop {} {\emph {\bibinfo {title} {{M}athematical
  {M}ethods for {P}hysicists}}},\ \bibinfo {edition} {3rd}\ ed.\ (\bibinfo
  {publisher} {Academic Press},\ \bibinfo {year} {1985})\BibitemShut {NoStop}%
\bibitem [{\citenamefont {de~Doncker-Kapenga}\ \emph
  {et~al.}(1983)\citenamefont {de~Doncker-Kapenga}, \citenamefont {{K}ahaner},
  \citenamefont {{P}iessens},\ and\ \citenamefont {\"{U}berhuber}}]{dedo83a}%
  \BibitemOpen
  \bibfield  {author} {\bibinfo {author} {\bibfnamefont {E.}~\bibnamefont
  {de~Doncker-Kapenga}}, \bibinfo {author} {\bibfnamefont {D.~K.}\ \bibnamefont
  {{K}ahaner}}, \bibinfo {author} {\bibfnamefont {R.}~\bibnamefont
  {{P}iessens}}, \ and\ \bibinfo {author} {\bibfnamefont {C.~W.}\ \bibnamefont
  {\"{U}berhuber}},\ }\href
  {http://www.ebook.de/de/product/17970317/e_de_doncker_kapenga_d_k_kahaner_r_piessens_c_w_ueberhuber_quadpack.html}
  {\emph {\bibinfo {title} {{QUADPACK}: {A S}ubroutine {P}ackage for
  {A}utomatic {I}ntegration}}}\ (\bibinfo  {publisher} {Springer},\ \bibinfo
  {year} {1983})\BibitemShut {NoStop}%
\bibitem [{\citenamefont {{T}ian}\ and\ \citenamefont
  {{C}hen}(2012)}]{tian12a}%
  \BibitemOpen
  \bibfield  {author} {\bibinfo {author} {\bibfnamefont {H.}~\bibnamefont
  {{T}ian}}\ and\ \bibinfo {author} {\bibfnamefont {G.}~\bibnamefont
  {{C}hen}},\ }\href {\doibase 10.1063/1.4767460} {\bibfield  {journal}
  {\bibinfo  {journal} {{J}. {C}hem. {P}hys.}\ }\textbf {\bibinfo {volume}
  {137}},\ \bibinfo {pages} {204114} (\bibinfo {year} {2012})}\BibitemShut
  {NoStop}%
\bibitem [{\citenamefont {Schr{\"o}ter}(2015)}]{schr15b}%
  \BibitemOpen
  \bibfield  {author} {\bibinfo {author} {\bibfnamefont {M.}~\bibnamefont
  {Schr{\"o}ter}},\ }\href {\doibase 10.1007/978-3-658-09282-5} {\emph
  {\bibinfo {title} {Dissipative Exciton Dynamics in Light-Harvesting
  Complexes}}}\ (\bibinfo  {publisher} {Springer},\ \bibinfo {year}
  {2015})\BibitemShut {NoStop}%
\bibitem [{\citenamefont {Feynman}\ and\ \citenamefont
  {Vernon}(1963)}]{feyn63a}%
  \BibitemOpen
  \bibfield  {author} {\bibinfo {author} {\bibfnamefont {R.~P.}\ \bibnamefont
  {Feynman}}\ and\ \bibinfo {author} {\bibfnamefont {F.~L.}\ \bibnamefont
  {Vernon}},\ }\href {\doibase 10.1016/0003-4916(63)90068-X} {\bibfield
  {journal} {\bibinfo  {journal} {Ann. Phys}\ }\textbf {\bibinfo {volume}
  {24}},\ \bibinfo {pages} {118} (\bibinfo {year} {1963})}\BibitemShut
  {NoStop}%
\bibitem [{\citenamefont {Xu}\ \emph {et~al.}(2005)\citenamefont {Xu},
  \citenamefont {Cui}, \citenamefont {Li}, \citenamefont {Mo},\ and\
  \citenamefont {Yan}}]{xu05a}%
  \BibitemOpen
  \bibfield  {author} {\bibinfo {author} {\bibfnamefont {R.-X.}\ \bibnamefont
  {Xu}}, \bibinfo {author} {\bibfnamefont {P.}~\bibnamefont {Cui}}, \bibinfo
  {author} {\bibfnamefont {X.-Q.}\ \bibnamefont {Li}}, \bibinfo {author}
  {\bibfnamefont {Y.}~\bibnamefont {Mo}}, \ and\ \bibinfo {author}
  {\bibfnamefont {Y.}~\bibnamefont {Yan}},\ }\href {\doibase 10.1063/1.1850899}
  {\bibfield  {journal} {\bibinfo  {journal} {J. Chem. Phys.}\ }\textbf
  {\bibinfo {volume} {122}},\ \bibinfo {pages} {041103} (\bibinfo {year}
  {2005})}\BibitemShut {NoStop}%
\bibitem [{\citenamefont {{S}chr\"oder}, \citenamefont {{S}chreiber},\ and\
  \citenamefont {{K}leinekath\"ofer}(2007)}]{schr06c}%
  \BibitemOpen
  \bibfield  {author} {\bibinfo {author} {\bibfnamefont {M.}~\bibnamefont
  {{S}chr\"oder}}, \bibinfo {author} {\bibfnamefont {M.}~\bibnamefont
  {{S}chreiber}}, \ and\ \bibinfo {author} {\bibfnamefont {U.}~\bibnamefont
  {{K}leinekath\"ofer}},\ }\href {\doibase 10.1063/1.2538754} {\bibfield
  {journal} {\bibinfo  {journal} {{J}. {C}hem. {P}hys.}\ }\textbf {\bibinfo
  {volume} {126}},\ \bibinfo {pages} {114102} (\bibinfo {year}
  {2007})}\BibitemShut {NoStop}%
\bibitem [{\citenamefont {{C}hen}\ \emph {et~al.}(2009)\citenamefont {{C}hen},
  \citenamefont {{Z}heng}, \citenamefont {{S}hi},\ and\ \citenamefont
  {{Y}an}}]{chen09b}%
  \BibitemOpen
  \bibfield  {author} {\bibinfo {author} {\bibfnamefont {L.}~\bibnamefont
  {{C}hen}}, \bibinfo {author} {\bibfnamefont {R.}~\bibnamefont {{Z}heng}},
  \bibinfo {author} {\bibfnamefont {Q.}~\bibnamefont {{S}hi}}, \ and\ \bibinfo
  {author} {\bibfnamefont {Y.}~\bibnamefont {{Y}an}},\ }\href {\doibase
  10.1063/1.3213013} {\bibfield  {journal} {\bibinfo  {journal} {{J}. {C}hem.
  {P}hys.}\ }\textbf {\bibinfo {volume} {131}},\ \bibinfo {eid} {094502}
  (\bibinfo {year} {2009})}\BibitemShut {NoStop}%
\bibitem [{\citenamefont {{I}shizaki}\ \emph {et~al.}(2010)\citenamefont
  {{I}shizaki}, \citenamefont {{C}alhoun}, \citenamefont {{S}chlau{C}ohen},\
  and\ \citenamefont {{F}leming}}]{ishi10a}%
  \BibitemOpen
  \bibfield  {author} {\bibinfo {author} {\bibfnamefont {A.}~\bibnamefont
  {{I}shizaki}}, \bibinfo {author} {\bibfnamefont {T.~R.}\ \bibnamefont
  {{C}alhoun}}, \bibinfo {author} {\bibfnamefont {G.~S.}\ \bibnamefont
  {{S}chlau{C}ohen}}, \ and\ \bibinfo {author} {\bibfnamefont {G.~R.}\
  \bibnamefont {{F}leming}},\ }\href {\doibase 10.1039/C003389H} {\bibfield
  {journal} {\bibinfo  {journal} {{P}hys. {C}hem. {C}hem. {P}hys.}\ }\textbf
  {\bibinfo {volume} {12}},\ \bibinfo {pages} {7319} (\bibinfo {year}
  {2010})}\BibitemShut {NoStop}%
\bibitem [{\citenamefont {{I}shizaki}\ and\ \citenamefont
  {{F}leming}(2011)}]{ishi11a}%
  \BibitemOpen
  \bibfield  {author} {\bibinfo {author} {\bibfnamefont {A.}~\bibnamefont
  {{I}shizaki}}\ and\ \bibinfo {author} {\bibfnamefont {G.~R.}\ \bibnamefont
  {{F}leming}},\ }\href {\doibase 10.1021/jp112406h} {\bibfield  {journal}
  {\bibinfo  {journal} {{J}. {P}hys. {C}hem. {B}}\ }\textbf {\bibinfo {volume}
  {115}},\ \bibinfo {pages} {6227} (\bibinfo {year} {2011})}\BibitemShut
  {NoStop}%
\bibitem [{\citenamefont {Aghtar}\ \emph {et~al.}(2012)\citenamefont {Aghtar},
  \citenamefont {Liebers}, \citenamefont {Str{\"{u}}mpfer}, \citenamefont
  {Schulten},\ and\ \citenamefont {Kleinekath{\"{o}}fer}}]{aght12a}%
  \BibitemOpen
  \bibfield  {author} {\bibinfo {author} {\bibfnamefont {M.}~\bibnamefont
  {Aghtar}}, \bibinfo {author} {\bibfnamefont {J.}~\bibnamefont {Liebers}},
  \bibinfo {author} {\bibfnamefont {J.}~\bibnamefont {Str{\"{u}}mpfer}},
  \bibinfo {author} {\bibfnamefont {K.}~\bibnamefont {Schulten}}, \ and\
  \bibinfo {author} {\bibfnamefont {U.}~\bibnamefont {Kleinekath{\"{o}}fer}},\
  }\href {\doibase 10.1063/1.4723669} {\bibfield  {journal} {\bibinfo
  {journal} {{J}. {C}hem. {P}hys.}\ }\textbf {\bibinfo {volume} {136}},\
  \bibinfo {pages} {214101} (\bibinfo {year} {2012})}\BibitemShut {NoStop}%
\bibitem [{\citenamefont {Hu}, \citenamefont {Gu},\ and\ \citenamefont
  {Franco}(2018)}]{hu18a}%
  \BibitemOpen
  \bibfield  {author} {\bibinfo {author} {\bibfnamefont {W.}~\bibnamefont
  {Hu}}, \bibinfo {author} {\bibfnamefont {B.}~\bibnamefont {Gu}}, \ and\
  \bibinfo {author} {\bibfnamefont {I.}~\bibnamefont {Franco}},\ }\href
  {\doibase 10.1063/1.5004578} {\bibfield  {journal} {\bibinfo  {journal} {J.
  Chem. Phys.}\ }\textbf {\bibinfo {volume} {148}},\ \bibinfo {pages} {134304}
  (\bibinfo {year} {2018})}\BibitemShut {NoStop}%
\bibitem [{\citenamefont {Wang}\ \emph {et~al.}(2016)\citenamefont {Wang},
  \citenamefont {Chen}, \citenamefont {Zhou},\ and\ \citenamefont
  {Zhao}}]{wang16b}%
  \BibitemOpen
  \bibfield  {author} {\bibinfo {author} {\bibfnamefont {L.}~\bibnamefont
  {Wang}}, \bibinfo {author} {\bibfnamefont {L.}~\bibnamefont {Chen}}, \bibinfo
  {author} {\bibfnamefont {N.}~\bibnamefont {Zhou}}, \ and\ \bibinfo {author}
  {\bibfnamefont {Y.}~\bibnamefont {Zhao}},\ }\href {\doibase
  10.1063/1.4939144} {\bibfield  {journal} {\bibinfo  {journal} {J. Chem.
  Phys.}\ }\textbf {\bibinfo {volume} {144}},\ \bibinfo {pages} {024101}
  (\bibinfo {year} {2016})}\BibitemShut {NoStop}%
\bibitem [{\citenamefont {Sun}\ \emph {et~al.}(2016)\citenamefont {Sun},
  \citenamefont {Fujihashi}, \citenamefont {Ishizaki},\ and\ \citenamefont
  {Zhao}}]{sun16b}%
  \BibitemOpen
  \bibfield  {author} {\bibinfo {author} {\bibfnamefont {K.-W.}\ \bibnamefont
  {Sun}}, \bibinfo {author} {\bibfnamefont {Y.}~\bibnamefont {Fujihashi}},
  \bibinfo {author} {\bibfnamefont {A.}~\bibnamefont {Ishizaki}}, \ and\
  \bibinfo {author} {\bibfnamefont {Y.}~\bibnamefont {Zhao}},\ }\href {\doibase
  10.1063/1.4950888} {\bibfield  {journal} {\bibinfo  {journal} {J. Chem.
  Phys.}\ }\textbf {\bibinfo {volume} {144}},\ \bibinfo {pages} {204106}
  (\bibinfo {year} {2016})}\BibitemShut {NoStop}%
\bibitem [{\citenamefont {{C}oifman}, \citenamefont {{R}okhlin},\ and\
  \citenamefont {{W}andzura}(1993)}]{coif93a}%
  \BibitemOpen
  \bibfield  {author} {\bibinfo {author} {\bibfnamefont {R.}~\bibnamefont
  {{C}oifman}}, \bibinfo {author} {\bibfnamefont {V.}~\bibnamefont
  {{R}okhlin}}, \ and\ \bibinfo {author} {\bibfnamefont {S.}~\bibnamefont
  {{W}andzura}},\ }\href {\doibase 10.1109/74.250128} {\bibfield  {journal}
  {\bibinfo  {journal} {IEEE Trans. Antennas Propag.}\ }\textbf {\bibinfo
  {volume} {35}},\ \bibinfo {pages} {7} (\bibinfo {year} {1993})}\BibitemShut
  {NoStop}%
\bibitem [{\citenamefont {Landa}, \citenamefont {Tanushev},\ and\ \citenamefont
  {Tsai}(2011)}]{land11a}%
  \BibitemOpen
  \bibfield  {author} {\bibinfo {author} {\bibfnamefont {Y.}~\bibnamefont
  {Landa}}, \bibinfo {author} {\bibfnamefont {N.~M.}\ \bibnamefont {Tanushev}},
  \ and\ \bibinfo {author} {\bibfnamefont {R.}~\bibnamefont {Tsai}},\ }\href
  {\doibase 10.4310/cms.2011.v9.n3.a11} {\bibfield  {journal} {\bibinfo
  {journal} {Commun. Math. Sci.}\ }\textbf {\bibinfo {volume} {9}},\ \bibinfo
  {pages} {903} (\bibinfo {year} {2011})}\BibitemShut {NoStop}%
\bibitem [{\citenamefont {{A}dolphs}\ and\ \citenamefont
  {{R}enger}(2006)}]{adol06a}%
  \BibitemOpen
  \bibfield  {author} {\bibinfo {author} {\bibfnamefont {J.}~\bibnamefont
  {{A}dolphs}}\ and\ \bibinfo {author} {\bibfnamefont {T.}~\bibnamefont
  {{R}enger}},\ }\href {\doibase 10.1529/biophysj.105.079483} {\bibfield
  {journal} {\bibinfo  {journal} {{B}iophys. {J}.}\ }\textbf {\bibinfo {volume}
  {91}},\ \bibinfo {pages} {2778} (\bibinfo {year} {2006})}\BibitemShut
  {NoStop}%
\bibitem [{\citenamefont {{K}ell}\ \emph {et~al.}(2013)\citenamefont {{K}ell},
  \citenamefont {{F}eng}, \citenamefont {{R}eppert},\ and\ \citenamefont
  {{J}ankowiak}}]{kell13a}%
  \BibitemOpen
  \bibfield  {author} {\bibinfo {author} {\bibfnamefont {A.}~\bibnamefont
  {{K}ell}}, \bibinfo {author} {\bibfnamefont {X.}~\bibnamefont {{F}eng}},
  \bibinfo {author} {\bibfnamefont {M.}~\bibnamefont {{R}eppert}}, \ and\
  \bibinfo {author} {\bibfnamefont {R.}~\bibnamefont {{J}ankowiak}},\ }\href
  {\doibase 10.1021/jp405094p} {\bibfield  {journal} {\bibinfo  {journal} {{J}.
  {P}hys. {C}hem. {B}}\ }\textbf {\bibinfo {volume} {117}},\ \bibinfo {pages}
  {7317} (\bibinfo {year} {2013})}\BibitemShut {NoStop}%
\bibitem [{\citenamefont {{K}reisbeck}, \citenamefont {{K}ramer},\ and\
  \citenamefont {Aspuru-Guzik}(2014)}]{krei14a}%
  \BibitemOpen
  \bibfield  {author} {\bibinfo {author} {\bibfnamefont {C.}~\bibnamefont
  {{K}reisbeck}}, \bibinfo {author} {\bibfnamefont {T.}~\bibnamefont
  {{K}ramer}}, \ and\ \bibinfo {author} {\bibfnamefont {A.}~\bibnamefont
  {Aspuru-Guzik}},\ }\href {\doibase 10.1021/ct500629s} {\bibfield  {journal}
  {\bibinfo  {journal} {{J}. {C}hem. {T}heory {C}omput.}\ }\textbf {\bibinfo
  {volume} {10}},\ \bibinfo {pages} {4045} (\bibinfo {year}
  {2014})}\BibitemShut {NoStop}%
\end{thebibliography}
%

\end{document}